%% file: main_text.tex
\documentclass[aos,preprint]{imsart}

\RequirePackage[OT1]{fontenc}
\RequirePackage{amsmath,amssymb,amsthm}
\RequirePackage[numbers]{natbib}
\RequirePackage[colorlinks,citecolor=blue,urlcolor=blue]{hyperref}
\usepackage{graphicx}

\startlocaldefs
\numberwithin{equation}{section}
\theoremstyle{plain}

\endlocaldefs

\theoremstyle{plain}
\long\def\comment#1{}

\theoremstyle{definition}

\numberwithin{definition}{section}

\numberwithin{remark}{section}

\renewcommand{\qed}{\hfill{\tiny \ensuremath{\blacksquare} }}%

\newcommand{\Ep}{{\mathrm{E}}}

\newcommand{\M}{\mathrm{M}}

\newcommand{\citeasnoun}{\cite}

\usepackage{multicol}
\usepackage{subcaption}
\usepackage{ragged2e} 
\usepackage{soul}

\begin{document}

\begin{frontmatter}
\title{
Closing the U.S. gender wage gap requires understanding its heterogeneity\thanksref{T1}  
}
\runtitle{Gender Wage Gap}
\thankstext{T1}{This version: \today}

\begin{aug}
\author{\fnms{Philipp} \snm{Bach}\thanksref{m1}\ead[label=e1]{} }
\and
\author{\fnms{Victor} \snm{Chernozhukov}\thanksref{m2}\ead[label=e2]{} \thanksref{T2}}
\and
\author{\fnms{Martin} \snm{Spindler}\thanksref{m1}\ead[label=e1]{}}

\thankstext{T2}{Corresponding author. The R-code and supplementary material will be provided upon request and published online at \url{https://www.bwl.uni-hamburg.de/en/statistik/forschung/software-und-daten.html}}

\affiliation{University of Hamburg\thanksmark{m1}, MIT\thanksmark{m2} and University of Hamburg\thanksmark{m1}}

\address{Philipp Bach\\
University of Hamburg\\
Hamburg Business School\\
Moorweidenstr. 18\\
20148 Hamburg\\
Germany\\
E-mail: philipp.bach@uni-hamburg.de}

\address{Victor Chernozhukov\\
Massachusetts Institute of Technology\\
Economics Department\\
USA\\
E-mail: vchern@mit.edu}

\address{Martin Spindler\\
University of Hamburg\\
Hamburg Business School\\
Moorweidenstr. 18\\
20148 Hamburg\\
Germany\\
E-mail: martin.spindler@uni-hamburg.de}
\end{aug}

\begin{abstract}
Abstract.
 In 2016, the majority of full-time employed women in the U.S. earned significantly less than comparable men. The extent to which women were affected by gender inequality in earnings, however, depended greatly on socio-economic characteristics, such as marital status or educational attainment. In this paper, we analyze data from the 2016 American Community Survey using a high-dimensional wage regression and applying double lasso to quantify heterogeneity in the gender wage gap. We found that the gap varied substantially across women and was driven primarily by marital status, having children at home, race, occupation, industry, and educational attainment. We recommend that policy makers use these insights to design policies that will reduce discrimination and unequal pay more effectively.
\end{abstract}



\end{frontmatter}


As a measure of inequality between men and women, the gender wage gap has come to play an important role both in academic research and the public debate. Most studies that have attempted to quantify gender inequality in earnings to date have reported wage gap estimates based on comparisons of the average wages of male and female employees. Most women in the labor force, however, experience wage penalties that differ from these depending on individual characteristics, such as educational attainment and occupation. Understanding the heterogeneity in the gender wage gap is crucial to designing effective and efficient occupation- and industry-specific programs that can lessen gender inequality in earnings for specific groups of female employees, such as married women and mothers.

In this paper, we aim to provide insights into the heterogeneity in the gender wage gap in the United States and, in doing so, contribute to a more comprehensive understanding of gender inequality in income. The extent to which the gender wage gap differs across women has attracted public attention and the interest of policy makers. While numerous policy reports and media articles have attempted to quantify heterogeneity in the wage gap, they have generally taken a simplistic approach based on comparing descriptive statistics across subgroups of people: They define these subgroups in terms of one characteristic only, such as region \citep{aauw2017}, race, ethnicity \citep{cnn2017,aauw2017,bls2017}, or occupation \citep{cap2015,wsj2016,vox2016}. Approaches such as this are likely to lead to flawed conclusions, however, because they neglect heterogeneity due to other variables. Moreover, statistical significance has only rarely been addressed although it becomes more and more important as the number of characteristics simultaneously being considered increases.

Studies in labor economics usually focus on the average wage gap between men and women, and perhaps control for a few hand-selected control variables \citep{blau2016}. In doing so, they fail to consider that the gender wage gap may vary with the control variables. We are aware of only several studies that examine heterogeneity in the wage gap and correlations with potential drivers of heterogeneity. For instance, in the study by \cite{goldin2014}, variation in the wage gap by occupation is considered and in \cite{goldin2017} the results are compared across specific industries and for married and never-married women. To the best of our knowledge, our study is the first to model heterogeneity in the gender wage gap in terms of a large set of socio-economic variables using representative data for the U.S. We estimated the gender wage gap for each full-time employed woman in the sample and illustrate the distribution of the wage gaps in quantile plots. To assess sources of heterogeneity, we estimated the effects of a rich set of potential determinants and report their joint statistical significance. 

We analyzed data from the 2016 American Community Survey (ACS), which is an annual survey of a representative sample of 1\% of the U.S. population. Participation in the survey is mandatory. We restricted our analysis to full-time and year-round employees and stratified the data according to participants’ educational attainment. The 2016 ACS collected data on a large number of socio-economic variables at the individual and household levels that we were able to use to model heterogeneity. We included information on marital status, having children at home (i.e., at least one biological, adopted or stepchild 18 years of age or younger), race, ethnicity (i.e., Hispanic origin), English language ability, geographic information (i.e., U.S. census region and metropolitan statistical area), veteran status, labor market characteristics (i.e., industry, occupation, hours worked), and the classic human capital variables (i.e., labor market experience and years of education). For people with a bachelor’s degree, we also included information on their college major. Our final data set comprised 642,229 observations, including 288,095 individuals who had attained a bachelor’s degree or higher (called “bachelor’s degree data” in the following) and 354,134 individuals with lower educational attainment, i.e., at most a high school diploma, GED or equivalent (called “high school degree data”). Further information on the composition of the sample, detailed descriptive statistics, and a description of our model and methodology are provided in the supplementary material available online.

Applying the traditional approach from the literature on labor economics, i.e., an Oaxaca-Blinder decomposition, to the 2016 ACS data results in a wage gap estimate of 17\% for people with a high school diploma or lower, and 14\% for people with a bachelor’s degree or higher (controlling for individual characteristics). However, the results of our analysis suggest that wage gaps of this magnitude are experienced only by a small proportion of women. To study heterogeneity in the wage gap, we allowed the gap to vary according to the socio-economic characteristics of the survey participants by including two-way interactions between the available variables. As doing so led to a large number of regression coefficients, we used double selection \citep{BelloniChernozhukovHansen2011}  for the least absolute shrinkage and selection operator \citep{T1996} (double lasso)  to estimate the high-dimensional (log) wage regression. This approach allowed us to estimate the gender wage gap – or, more precisely, the relative loss in pay compared to a man with the same socio-economic characteristics – for each woman in the data set.  We additionally report the ordinary least squares (OLS) results to allow for comparison.

\begin{figure}[!ht]
	 \begin{minipage}{\linewidth}
\centering
  \subcaption{Quantiles of effects with corresponding confidence bands,  DOUBLE LASSO.}	
	\label{qplots1}
	\end{minipage}
  \begin{minipage}{0.4\linewidth}
   \includegraphics[scale=0.3]{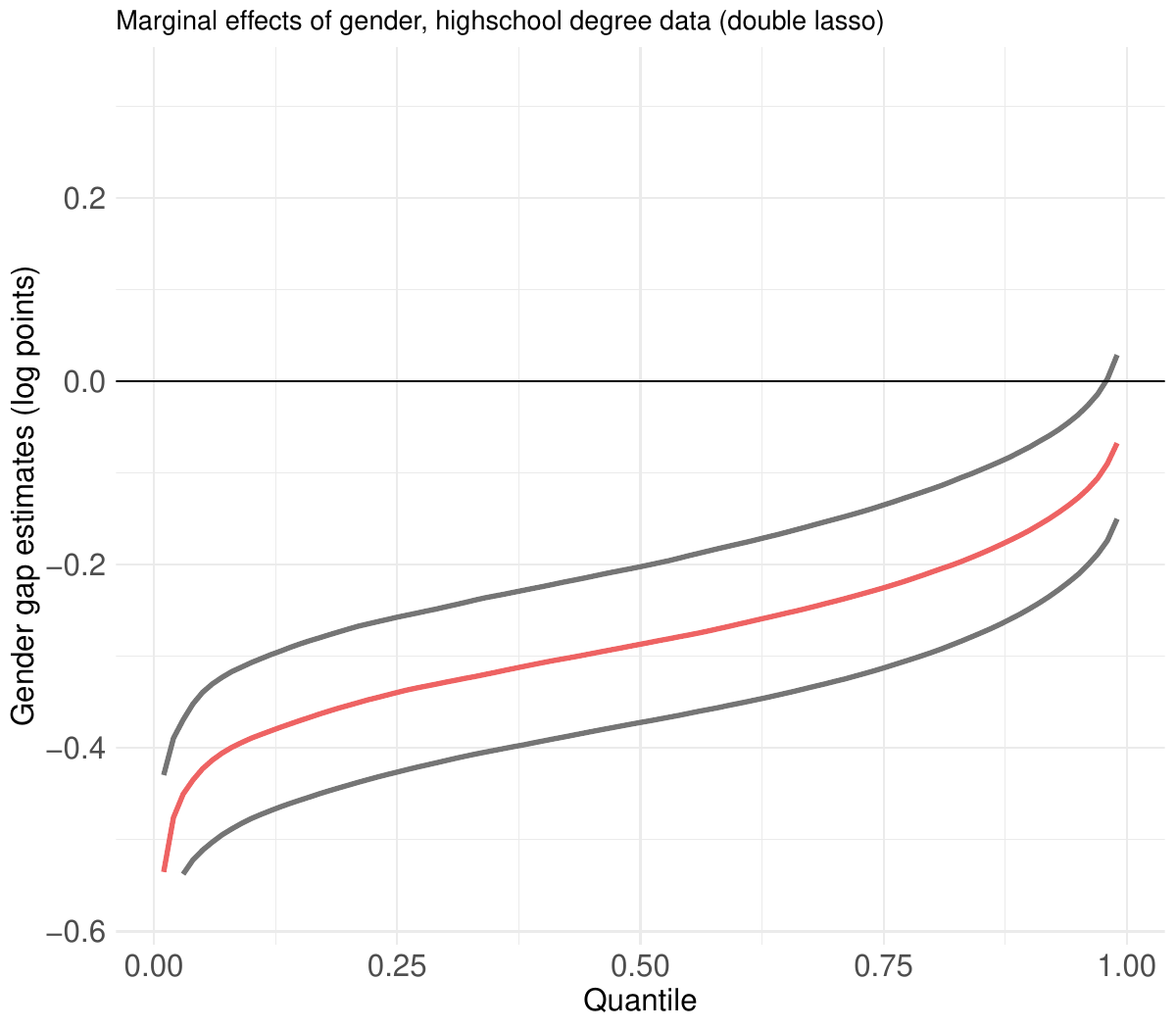} 
   \end{minipage}
  \hfill
  \begin{minipage}{0.49\linewidth}
    \includegraphics[scale=0.3]{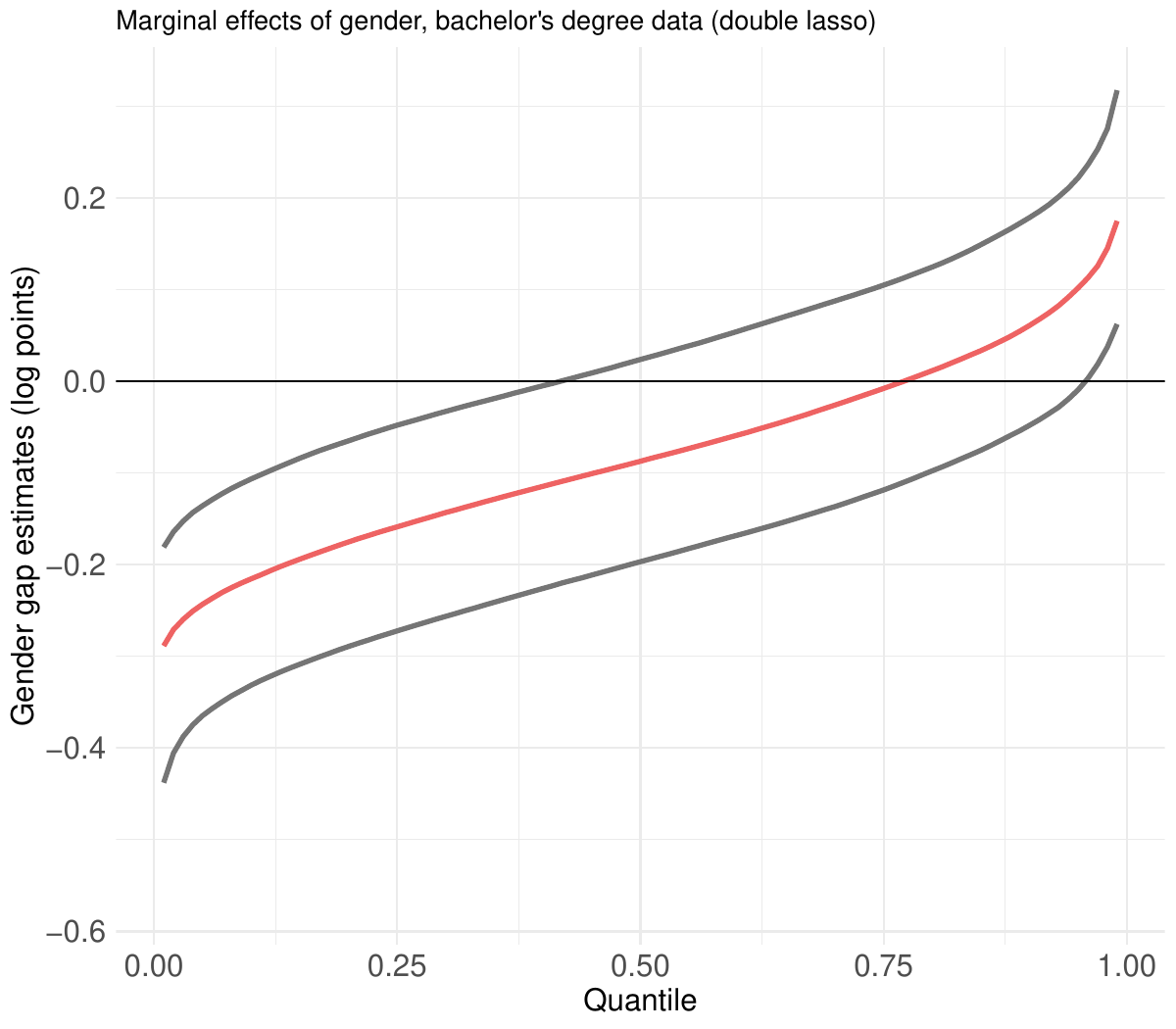}
  \end{minipage} 
		 \begin{minipage}{\linewidth}
\centering
  \subcaption{Quantiles of effects with corresponding confidence bands, OLS.}	
\end{minipage} 
	\begin{minipage}{0.49\linewidth}
   \includegraphics[scale=0.3]{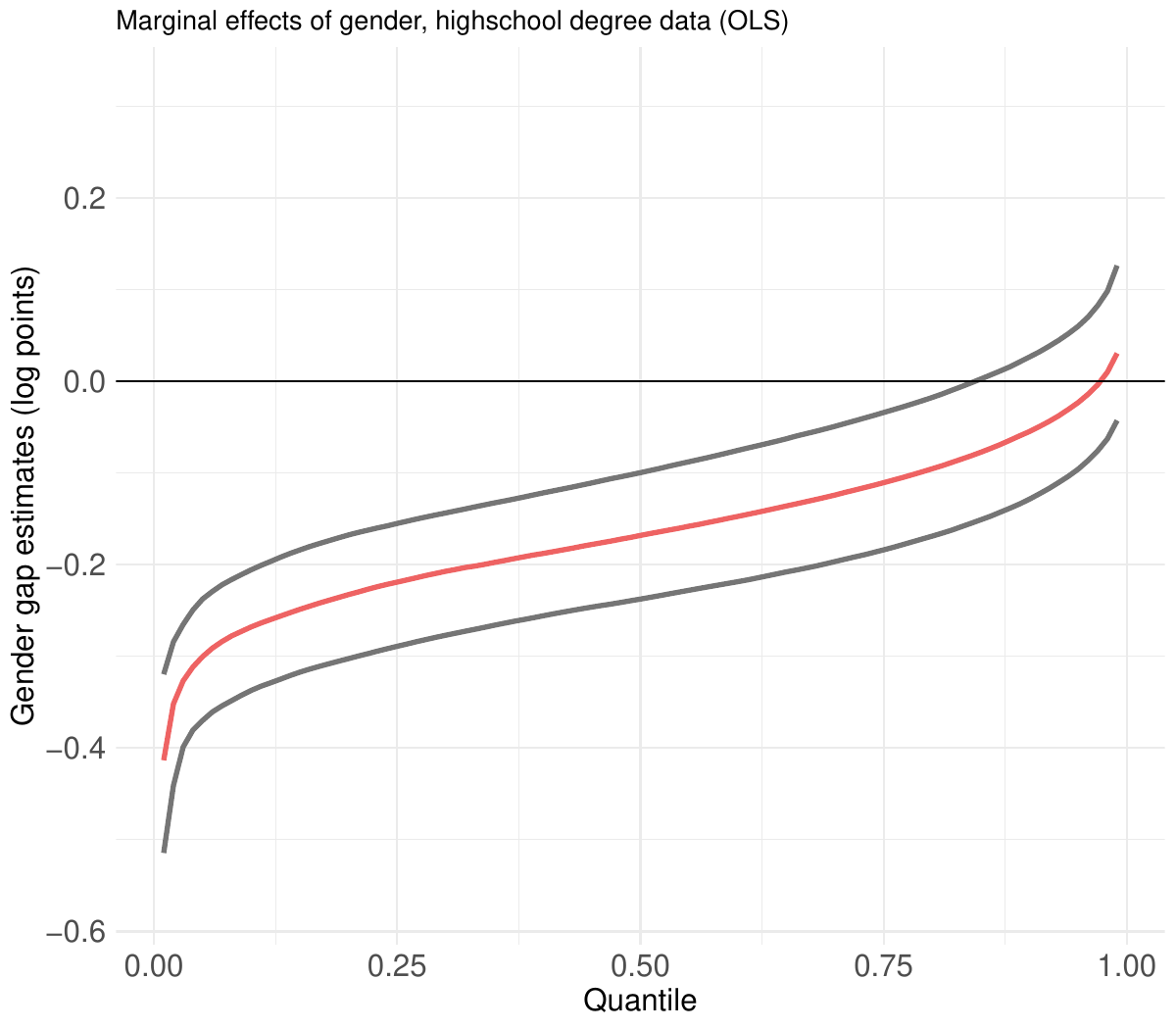} 
  \end{minipage}
  \hfill
  \begin{minipage}{0.49\linewidth}
    \includegraphics[scale=0.3]{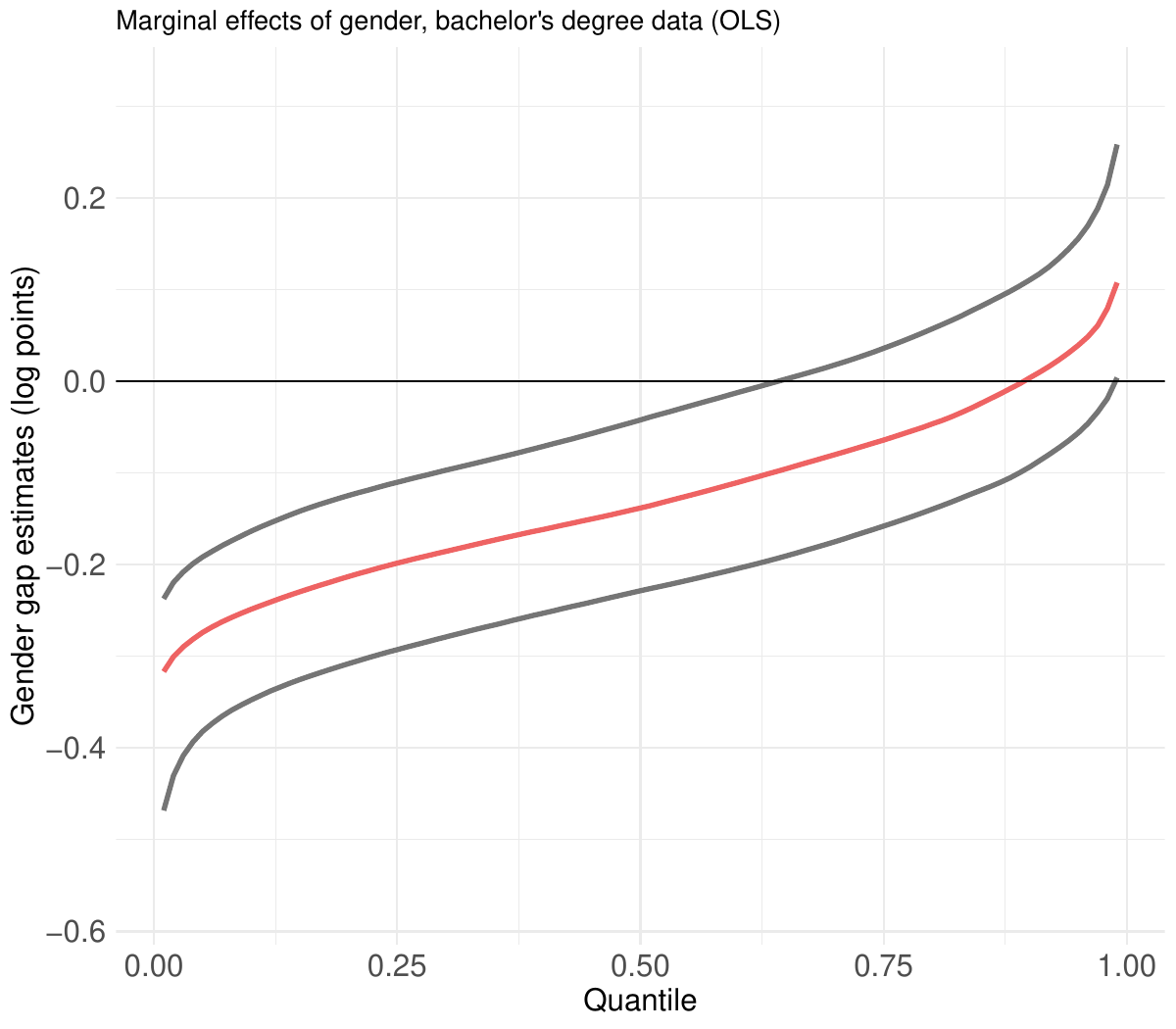}
  \end{minipage} 
  \caption{Quantiles of effects with simultaneous confidence bands.}
	\label{qplots}
\begin{justify} \footnotesize
The plots show the quantiles of the individual gender wage gap estimates as computed for
all women in the educational attainment subgroups of the ACS 2016 data together with simultaneous 0.95 confidence bands (gray
lines). Estimates in Panel (a) are obtained from a high-dimensional wage regression using the double 
lasso estimator, with log weekly wages as the dependent variable. Plots on the left refer to the high
school degree subgroup and plots on the right to the bachelor’s degree subgroup. In addition, ordinary least squares results are provided in Panel (b) for reasons of comparison.
\end{justify}
	\end{figure}

Figure \ref{qplots} provides quantile plots of the estimated wage gaps and illustrates that these were highly heterogeneous. Rather than affecting all women to the same extent, gender inequality in wages consisted of a range of wage penalties that differed greatly from woman to woman. For most women, the estimated gap deviated from the abovementioned estimates, derived from traditional analysis, of 17\% for people with a high school diploma or lower educational attainment and 14\% for people with a bachelor’s degree or higher. Patterns of heterogeneity varied substantially across the two samples, with gender wage inequality being more prevalent and more severe among women with lower educational attainment. Whereas more than 90\% of female employees with a high school degree or lower earned significantly less than their male counterparts, only 40\% of female employees with a bachelor’s degree or higher experienced a significant wage penalty according to the double lasso results in Figure \ref{qplots1}. Moreover, at any given quantile, the wage gap was larger for women who did not have a college degree. The median of the estimated wage gaps was around 9\% (non-significant) for women with postsecondary education. In contrast, a wage gap of at least 29\% was experienced by half of the women with lower educational attainment. Interestingly, there was evidence of a reversal of the gender wage gap for a small share of women with a college degree, i.e., 4\% of the full-time and full-year employees with postsecondary education earned significantly more than comparable men according to our double lasso results.

\begin{figure}[t]
   \includegraphics[scale=0.25]{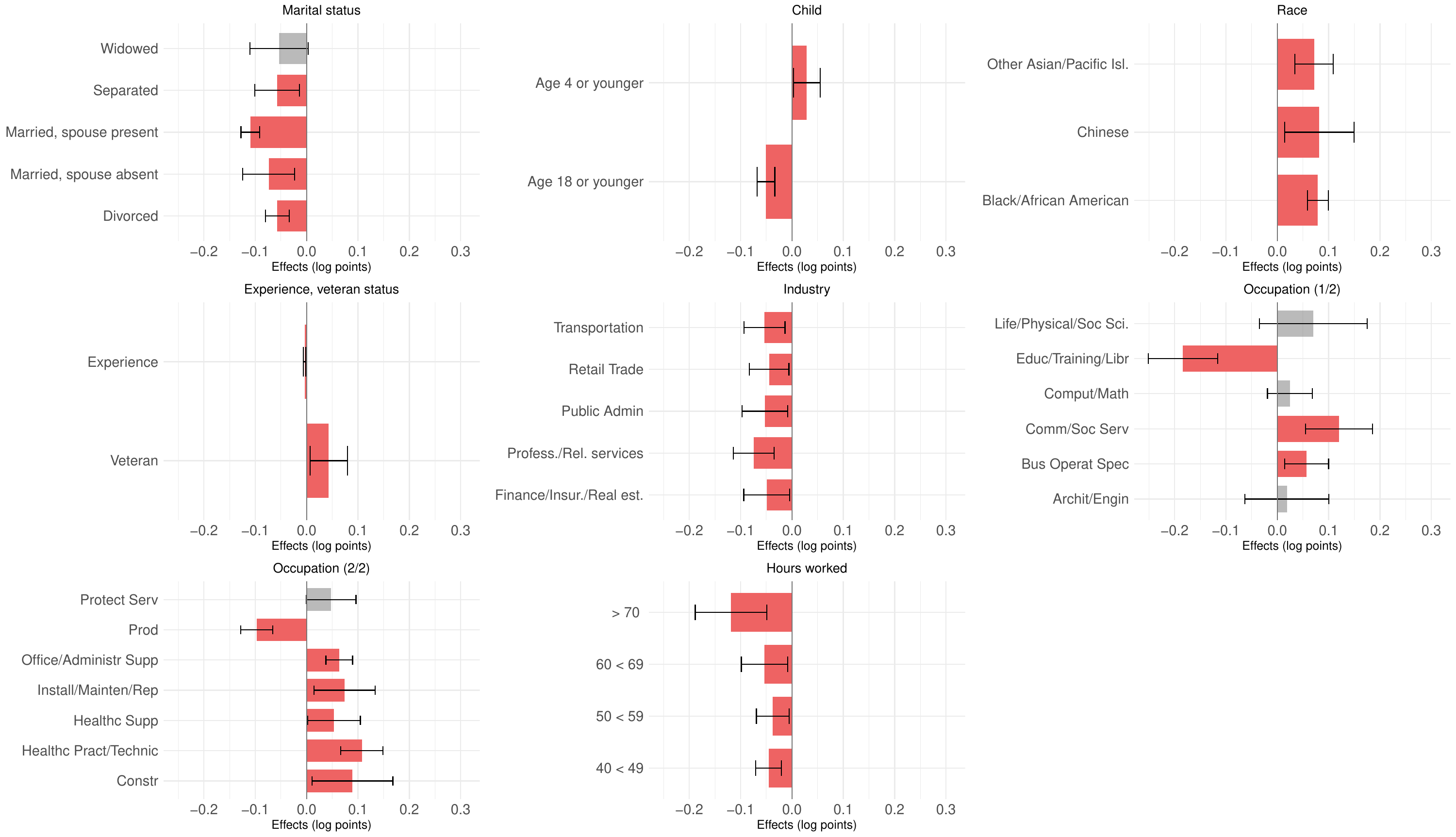}
	\caption{Effects of selected variables on the gender wage gap, high school degree subgroup.}
	\label{effectshighschool}
	\begin{justify} \footnotesize
The plot presents selected effects of socio-economic variables on the magnitude of the
gender wage gap with joint 0.95 confidence bands (black bounds). Effects indicate
significant changes in the wage gap compared to the baseline category indicated by
the vertical gray line. Baseline categories are: never married; no biological, adopted or stepchildren at home aged 4 or younger; no biological, adopted or stepchildren at home aged 18 or younger; White; not a veteran; wholesale trade (industry); management, business, science and arts (occupation); 35 to 40 hours work each week.
\end{justify}
		\end{figure}

\begin{figure}[t]
   \includegraphics[scale=0.25]{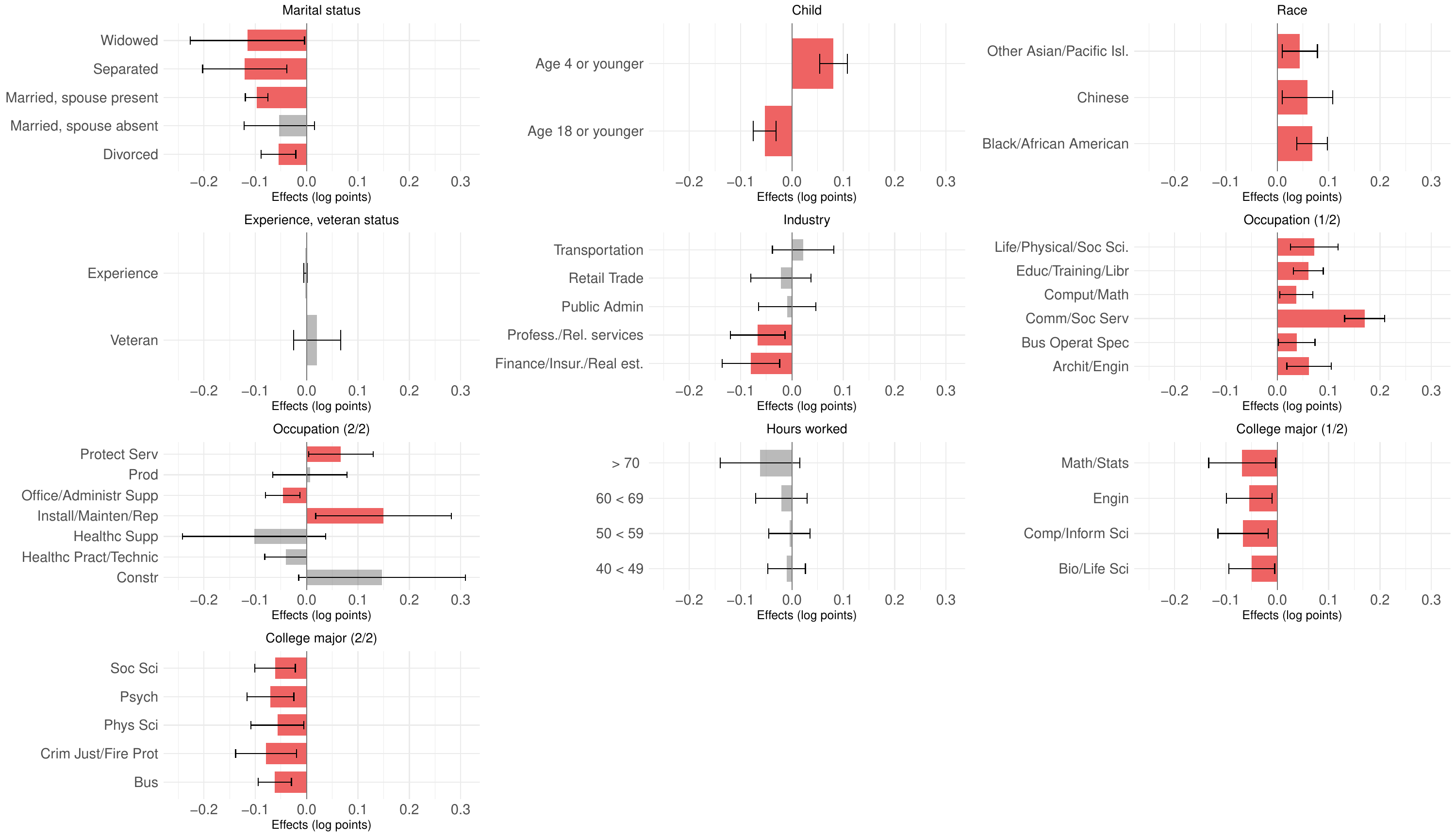}  
	\caption{Effects of selected variables on the gender wage gap, bachelor’s degree subgroup.}
		\label{effectsbachelor}
\begin{justify} \footnotesize
The plot presents selected effects of socio-economic variables on the magnitude of the
gender wage gap with joint 0.95 confidence bands (black bounds). Effects indicate
significant changes in the wage gap compared to the baseline category indicated
by the vertical gray line. Baseline categories are: never married; no biological, adopted or stepchildren at home aged 4 or younger; no biological, adopted or stepchildren at home aged 18 or younger; White; not a veteran; wholesale trade (industry); management, business, science and arts (occupation); 35 to 40 hours work each week; and education administration and teaching (college major).
\end{justify}
		\end{figure}

In the next step, we analyzed the drivers of heterogeneity in the wage gap.   Figures \ref{effectshighschool} and \ref{effectsbachelor} present selected estimated effects of discrete regressors on the gender wage gap compared to the baseline group (indicated by the gray vertical line). Negative changes are interpreted as an increase in the absolute value of the wage gap.


In both of the educational attainment subgroups, the gender wage gap showed similar patterns of heterogeneity for the variables marital status, having children at home and race. The effects associated with job-related variables, such as industry, occupation, and hours worked, differed in sign and magnitude across the two subgroups. A particularly large wage penalty was associated with marriage in both subgroups, with married women with spouse present experiencing a gap that, all other things being equal, was around nine to 12 percentage points larger than for women who had never been married. Moreover, our results point to a “motherhood penalty” \citep{siglerushton2007, waldfogel1998, angelov2016}  given that the wage gap was around five percentage points larger for women who had biological, adopted or stepchildren aged 18 or younger living with them. Apparently, the penalty for motherhood is rather time-persistent and does not directly reduce earnings of new mothers, at least for highly educated women. In this sample, mothers who had at least one child aged four or younger at home experienced a wage gap that was two percentage points smaller than for women without children, i.e., women who did not reside with a child aged 18 or younger. It is possible to observe a similar pattern with regard to having a young child at home in the high school subsample. However, the effect of having a young child at home is not sufficiently strong to lead to a comparable reversal, but rather reduces the magnitude of the wage gap for this subgroup. 

Interestingly, the wage gap was significantly smaller for races other than White in both samples. The gender gap differentials were more pronounced in the subgroup with lower educational attainment. Hence, the often reported variation of the gender gap according to race \citep{aauw2017} was robust to controlling for a large set of characteristics including education, experience, occupation, industry, and level of English language ability. The wage gap did not vary significantly across Hispanics and non-Hispanics.

The effects of the classic human capital variables “years of education” and “labor market experience” were small and non-significant in most cases. In the bachelor’s degree data, the gender wage gap varied according to college major. For 10 majors, including the natural sciences, social sciences and business, the wage gap was significantly larger than in the baseline major (i.e., the category ``education administration and teaching'').

Patterns of heterogeneity related to the work environment of full-time employees differed across the two samples, particularly the effects associated with occupation. This difference might be explained by differences in the temporal flexibility of jobs for employees who are highly qualified compared to employees with less education \citep{goldin2014, blau2016}. In concordance with such an explanation, the results of our analysis suggest that, among people with a high school degree or lower, the gender wage gap increased along with the number of working hours. Heterogeneity with respect to industry was found to be stronger for the high school degree subgroup. A common observation in both subgroups is that wage gaps were relatively large in the finance, insurance, and real estate industries, as well as in professional and related services.

In summary, our empirical analysis reveals that in 2016, most full-time employed women in the U.S. experienced a substantial wage penalty compared to observationally equivalent men. However, the extent to which women were affected by gender inequality in earnings differed greatly according to individual characteristics, including educational attainment, marital status, having children at home, race, and job-related characteristics such as occupation and industry. The commonly used average estimates of the gender wage gap can therefore be seen as a poor approximation of the wage penalty experienced by most women. By illustrating and quantifying heterogeneity in the wage gap, we hope to contribute to both the public and the academic discussion, and to provide information that policy makers can use to design more effective policies.

\clearpage
\footnotesize
\pagebreak
\bibliographystyle{imsart-number}
\bibliography{mybib}

\end{document}


\begin{frontmatter}
\title{\textit{Online Appendix} to \\ Closing the U.S. gender wage gap requires understanding its heterogeneity \thanksref{T1}}

\runtitle{Gender Wage Gap}
\thankstext{T1}{This version: \today}

\begin{aug}
\author{\fnms{Philipp} \snm{Bach}\thanksref{m1}\ead[label=e1]{} }
\and
\author{\fnms{Victor} \snm{Chernozhukov}\thanksref{m2}\ead[label=e2]{}\thanksref{T2}}
\and
\author{\fnms{Martin} \snm{Spindler}\thanksref{m1}\ead[label=e1]{}}

\thankstext{T2}{Corresponding author. The R-code will be provided online as supplementary material at \url{https://www.bwl.uni-hamburg.de/en/statistik/forschung/software-und-daten.html}.}


\affiliation{University of Hamburg\thanksmark{m1}, MIT\thanksmark{m2} and University of Hamburg\thanksmark{m1}}

\address{Philipp Bach\\
University of Hamburg\\
Hamburg Business School\\
Moorweidenstr. 18\\
20148 Hamburg\\
Germany\\
E-mail: philipp.bach@uni-hamburg.de}

\address{Victor Chernozhukov\\
Massachusetts Institute of Technology\\
Economics Department\\
USA\\
E-mail: vchern@mit.edu}

\address{Martin Spindler\\
University of Hamburg\\
Hamburg Business School\\
Moorweidenstr. 18\\
20148 Hamburg\\
Germany\\
E-mail: martin.spindler@uni-hamburg.de}

\end{aug}

\end{frontmatter}



\section{Supplementary Materials and Methods}

\subsection{Data description and replicability} \label{smsection1}

The 2016 American Community Survey (ACS) data used in the empirical analysis are provided by \cite{ipums} and can be extracted from the IPUMS-USA website (\url{https://usa.ipums.org/usa/}). The sample composition and the data analysis are fully reproducible. The R code and a guide for data extraction and preprocessing will be made available upon request and  be published online. 

The ACS provides a  representative 1\%-sample of the U.S. population. Participation in the survey is mandatory. A large number of socio-economic characteristics at the individual and household level are available, e.g., referring to education, industry, and occupation.  
We restricted attention to employed individuals working full time (35+ hours) and year-round, i.e., at least 50 weeks a year, to compare men and women with a similarly strong attachment to the labor force. Weekly earnings were computed as annual earnings divided by 52 (weeks). We focused on individuals aged 25 to 65 and discarded persons with income below the mandated federal minimum level of wages corresponding to an hourly wage of $\$7.25$ or - in terms of annual wage income - to $\$12,687.50$ according to our sample composition. As the federal minimum level has not been adjusted since 2009, we consider our exclusion rule as not restrictive. However, the rule was sufficient to exclude unrealistic weekly wages, e.g., wages corresponding to less than  $\$1$ per hour.  The final data set comprised $642,229$ individual observations and was stratified into two subgroups according to individuals' highest educational degree. The ``bachelor's degree data" comprised $288,095$ individuals with at least a bachelor's degree and the ``high school degree data" comprised $354,134$ observations with at most graduation from high school, GED or equivalent. 

\subsection{Variable construction}
In the empirical analysis, we used a set of 16 initial regressors to model heterogeneity in the gender wage gap. The variables are listed in Table \ref{listofvars} together with information on the baseline categories. Descriptive statistics are provided in Table \ref{summarystats}. 
The dependent variable in the wage regression is log weekly wage, i.e., wage gap estimates are reported in log scale throughout the paper. We modeled parenthood by including two binary variables. The first of these variables indicates that a person resided with one or more biological, adopted or stepchildren of age 18 or younger. The second variable takes on value one if a person lives in the same household with a biological, adopted or stepchild aged 4 or younger. We included both variables to analyze heterogeneity in the motherhood penalty in terms of the age of the child. We used the 14 major groups of the 1990 Census Bureau industry classification scheme available in the ACS (3-digit). Similarly, the Census Bureau provides a 2010 ACS classification of occupations (4-digit) that  are clustered in 26 major categories in the ACS. The variable ``hours worked'' is a categorical variable indicating the number of hours usually worked each week in the last 12 months. 
For the bachelor's degree subgroup, we additionally included the variable ``college major'' to account for individuals' educational background in more detail. 

To model  heterogeneity in the wage gap, we constructed all two-way interactions of the initial regressors and ended up with a high-dimensional setting with in total  $2{,}068$ (high school degree subgroup) and $4{,}382$ (bachelor's degree subgroup) regressors (categorical variables are transformed to level-wise dummies and variables with zero-variation are dropped). Of these regressors, $71$ (high school) and 106 (bachelor) refer to the initial set of characteristics $x_i$ and $1{,}997$ (high school degree) and $4{,}276$ (bachelor's degree) to the interacted regressors $z_i$ in regression Equation \ref{extendedwage} in Section \ref{methodology}.

\begin{table}[t]
\centering
\begin{tabular}{lrrrr}
  \hline 	  \\[0.8ex]
  Variable & Type &  Baseline Category  \\[0.8ex]
  \hline  \\[-0.8ex]
\multicolumn{3}{l}{\textit{Dependent Variable}} \\[0.8ex]
Log weekly wage & continuous &  \\
& & \\
\multicolumn{3}{l}{\textit{Independent Variables}} \\[0.8ex]
Female & binary &    \\ 
Marital status & 6 categories  & never married/single \\
Child age $\le$ 4 & binary & \\
(One or more biological, adopted or  & & \\
stepchildren at home aged 4 or younger) & & \\
Child age $\le$ 18 & binary & \\
(One or more biological, adopted or &  & \\
 stepchildren  at home aged 18 or younger) & & \\
Race & 4 categories &   White \\ 
Hispanic & binary &  \\
English language ability & 5 categories &  speaks only English \\ 
Experience (years) & continuous & \\ 
Experience squared & continuous & \\
Years of education & continuous & \\
Veteran status & binary  & \\ 
Industry & 14 categories  & wholesale trade\\
Occupation & 26 catergories  & management, business,  \\ 
 & & science, and arts\\
Hours worked each week & 5 categories & 35 to 40 hours \\
(usually worked in last 12 months) & & \\
College major  & 37 categories & education administration  \\
(Bachelor's degree data only) & & and teaching\\
Region (U.S. census) & 9 categories & New England division \\ 
MSA & binary &  \\ 
(metropolitan statistical area) & & \\
   \hline \\[0.8ex]
\end{tabular} 
\centering
 \caption{List of variables.} \label{listofvars}
\end{table} 

\clearpage 

\subsection{Summary statistics}

Table \ref{summarystats} provides summary statistics for a selection of variables available in the 2016 ACS data.  The descriptive statistics illustrate that wages were substantially higher for college graduates on average.  As expected, the individuals holding at least a bachelor degree were in education for a longer time and had less experience, on average. The shares of Hispanics and Blacks were lower in  the bachelor's degree subgroup, whereas the share of Chinese was higher. College graduates tended to live in metropolitan statistical areas more frequently and to work longer  hours, on average. Also the shares of persons who lived with their biological, adopted or stepchildren aged 18 or younger were higher in the bachelor's degree data. Similarly, the share of persons who resided with their biological, adopted or stepchildren aged four or younger was higher in the sample of college graduates. The patterns in terms of marital status differed across the educational subgroups. The share of married (with spouse present) persons was higher in the bachelor's degree data.

In both samples, average earnings of men exceeded those of women by far, both in terms of the mean (around 32\% for the high school and 49\% for the bachelor sample) and median weekly wage (33\% and 42\%). Although the corresponding unadjusted median wage gaps were slightly larger than recently reported by the U.S. Census Bureau\footnote{https://www.census.gov/data/tables/2016/demo/industry-occupation/acs-2016.html}, their magnitude was still comparable. The difference with respect to the unadjusted wage gap estimates probably arose because of different sample definitions, in particular due to the imposed requirement on year-round employees and the minimum wage criterion. 

\begin{figure}[t]
	\includegraphics[width = \linewidth]{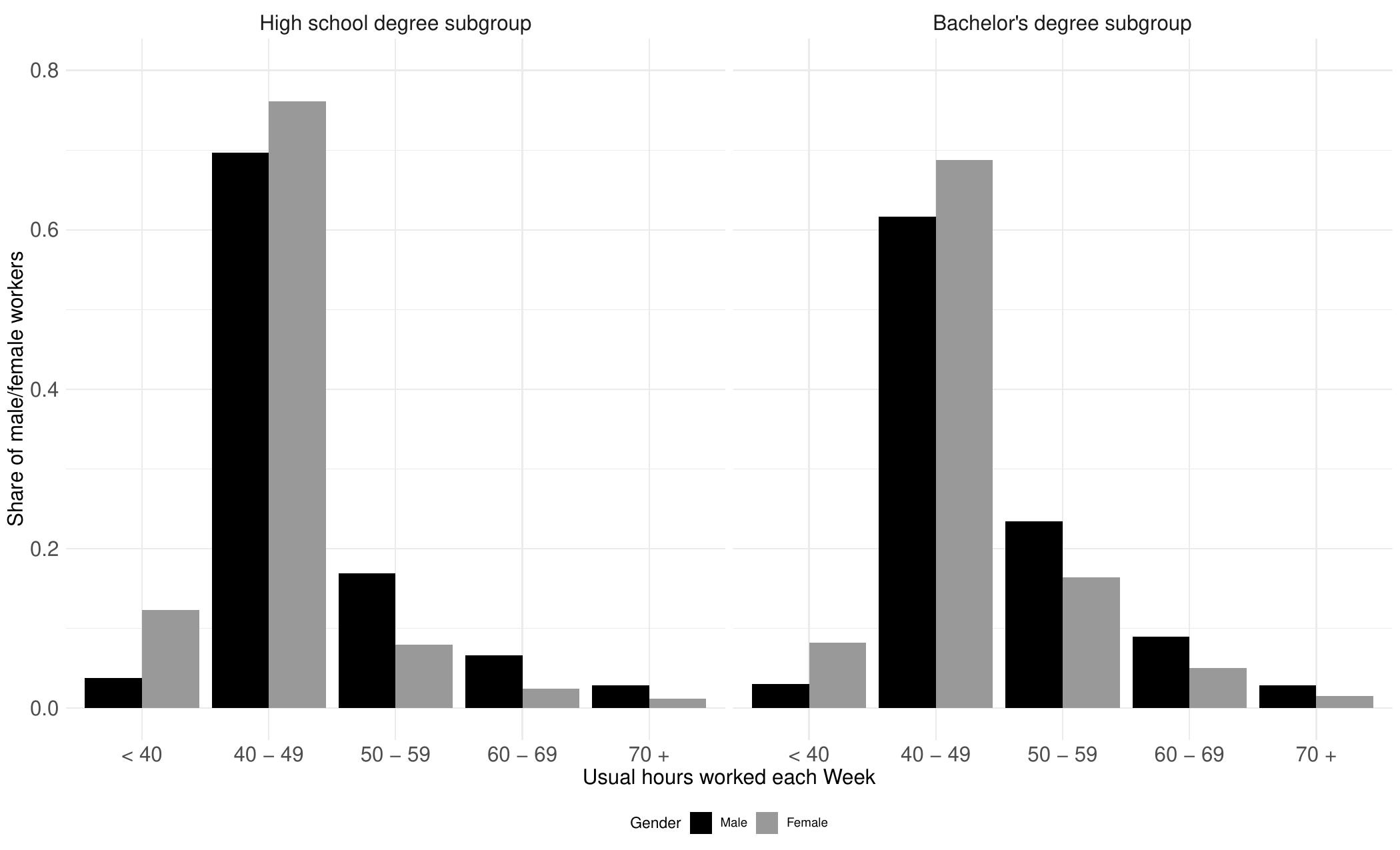}
\caption{Usual hours worked per week by gender.}
\label{hwplot}
\begin{justify}
The bar plot in Figure \ref{hwplot} illustrates the distribution of male (black bars) and female (gray bars) employees across categories of usual hours worked each week, separately for the two educational attainment subgroups. 
\end{justify}
	\end{figure}

An interesting descriptive finding was made with regard to the human capital characteristics ``years of education'' and ``experience''. The summary statistics for the high school degree data corresponded to the frequently mentioned ``reversal of the gender gap'' in terms of labor market characteristics. However, we could not confirm this observation for the sample of college graduates, probably due to selection into full-time employment. The gender gap in terms of years of education was virtually zero. Moreover, we observed that the gender gap in terms of hours worked was still considerable with men working for about 2.4 (bachelor's degree) and 3 (high school degree) hours each week longer than their female colleagues, on average. 
Figure \ref{hwplot} illustrates the fact  that the share of men in the group of employees who regularly work overtime is disproportionately large. 

\begin{table}[!thb]
\begin{tabular}{l rrrr }
  \hline && &&\\ 
Variable & 	\multicolumn{2}{r}{High school degree subgroup} & \multicolumn{2}{r}{Bachelor's degree subgroup}  \\[0.8ex]   
				& Men & Women & Men & Women  \\[0.8ex]  
  \hline && && \\[0.8ex]   
	Weekly wage (mean) & 1{,}098.95 & 833.83 & 2{,}244.29 & 1{,}508.86 \\ 
   & (863.34) & (618.87) & (1{,}996.49) & (1{,}240.98) \\ 
		& & & & \\
	  Weekly wage (median) & 923.08 & 692.31 & 1692.31 & 1192.31 \\ 
	& & & & \\
  	Single/never married & 0.21 & 0.19 & 0.19 & 0.24 \\ 
   & (0.16) & (0.16) & (0.15) & (0.18) \\ 
  Married, spouse present & 0.63 & 0.54 & 0.72 & 0.60 \\ 
   & (0.23) & (0.25) & (0.20) & (0.24) \\
	Child age $\le$ 4  & 0.12 & 0.08 & 0.16 & 0.13 \\ 
   & (0.11) & (0.07) & (0.14) & (0.11) \\ 
  Child age $\le$ 18 & 0.37 & 0.33 & 0.44 & 0.38 \\ 
   & (0.23) & (0.22) & (0.25) & (0.24) \\ 
 White & 0.86 & 0.81 & 0.84 & 0.81 \\ 
   & (0.12) & (0.16) & (0.14) & (0.16) \\ 
  Black & 0.10 & 0.15 & 0.05 & 0.09 \\ 
   & (0.09) & (0.12) & (0.05) & (0.08) \\ 
  Chinese & 0.01 & 0.01 & 0.03 & 0.03 \\ 
   & (0.01) & (0.01) & (0.03) & (0.03) \\ 
  Hispanic & 0.14 & 0.12 & 0.06 & 0.06 \\ 
   & (0.12) & (0.11) & (0.05) & (0.06) \\ 
 Experience (years) & 27.03 & 28.48 & 21.61 & 20.38 \\ 
   & (11.23) & (11.08) & (10.99) & (11.19) \\ 
  Years of education & 12.43 & 12.66 & 16.95 & 16.97 \\ 
    & (1.17) & (1.06) & (1.28) & (1.23) \\ 
  Hours worked (mean) & 44.93 & 41.90 & 46.20 & 43.79 \\ 
   & (8.56) & (6.34) & (8.65) & (7.34) \\ 
  Hours worked (median) & 40 & 40 & 43 & 40 \\ 
		& & & & \\
		  Veteran status & 0.11 & 0.01 & 0.07 & 0.02 \\ 
   & (0.31) & (0.12) & (0.25) & (0.12) \\ 
  MSA & 0.85 & 0.86 & 0.94 & 0.93 \\ 
   & (0.36) & (0.35) & (0.23) & (0.26) \\ 
		& & & & \\
  No. of observations & 207{,}549 & 146{,}585 & 154{,}833 & 133{,}262 \\ 
  (\%) & 58.61 & 41.39 & 53.74 & 46.26 \\ 
	\hline \\[0.8ex]
	 \end{tabular}
\caption{Summary statistics.}
\label{summarystats}
\begin{justify}
The table presents summary statistics for selected observable characteristics, separately for men and women in each educational attainment subgroup. 
Mean values are presented for selected observable characteristics. The median of the income variable and the median of the number of hours worked each week are presented in each subgroup, as well. Standard deviation in parentheses. The summary statistics are calculated for the subsample of the American Community Survey 2016 according to the sample definition presented in  Section \ref{smsection1} in the supplementary material. 
\end{justify}
\end{table}

\clearpage

\subsection{Details on preliminary results from a traditional Oaxaca-Blinder decomposition}

\begin{figure}[t]
	\begin{minipage}{0.45\linewidth}
   \includegraphics[scale=0.3]{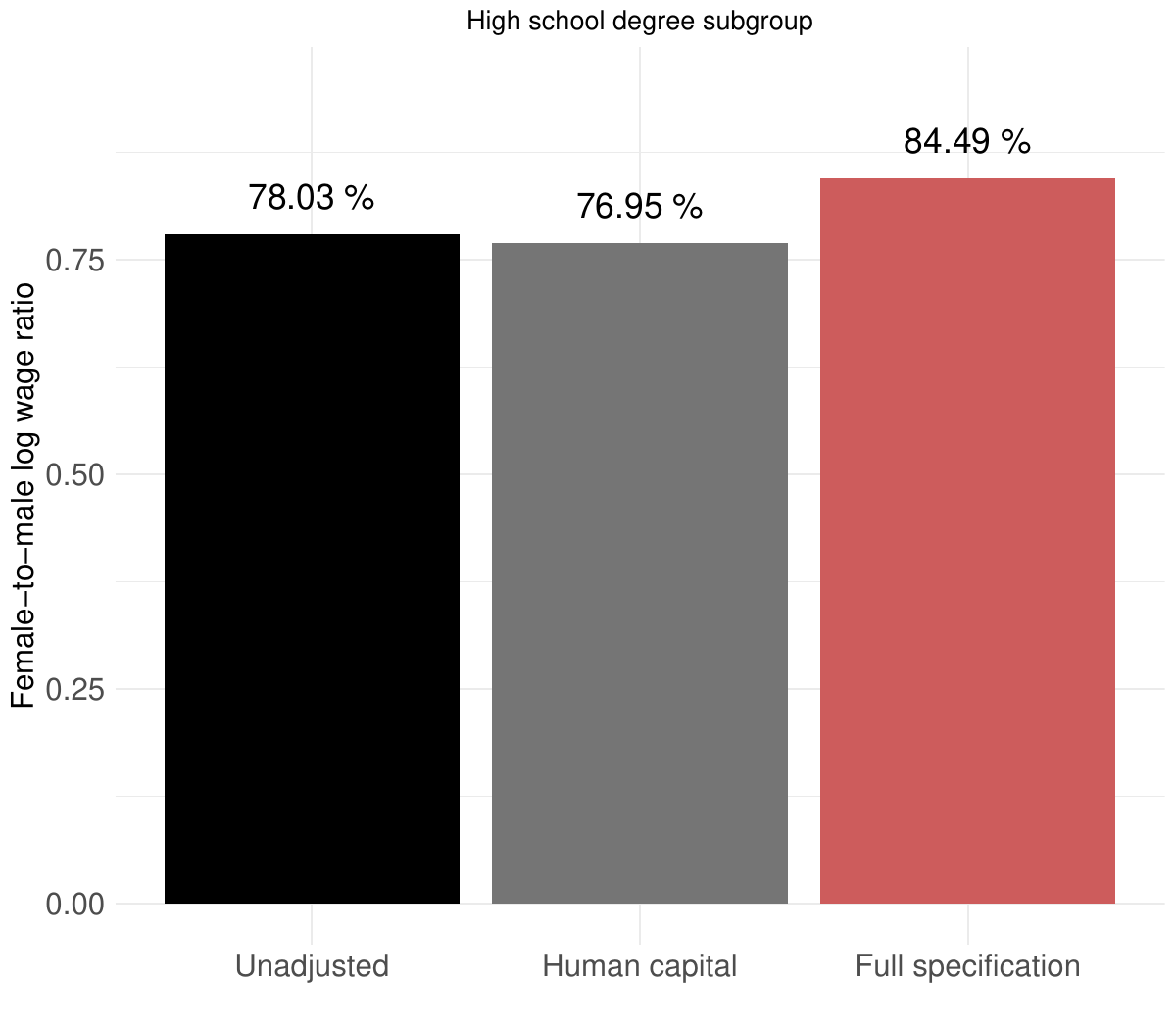}
  \end{minipage}
	 \hfill
		\begin{minipage}{0.45\linewidth}
   \includegraphics[scale=0.3]{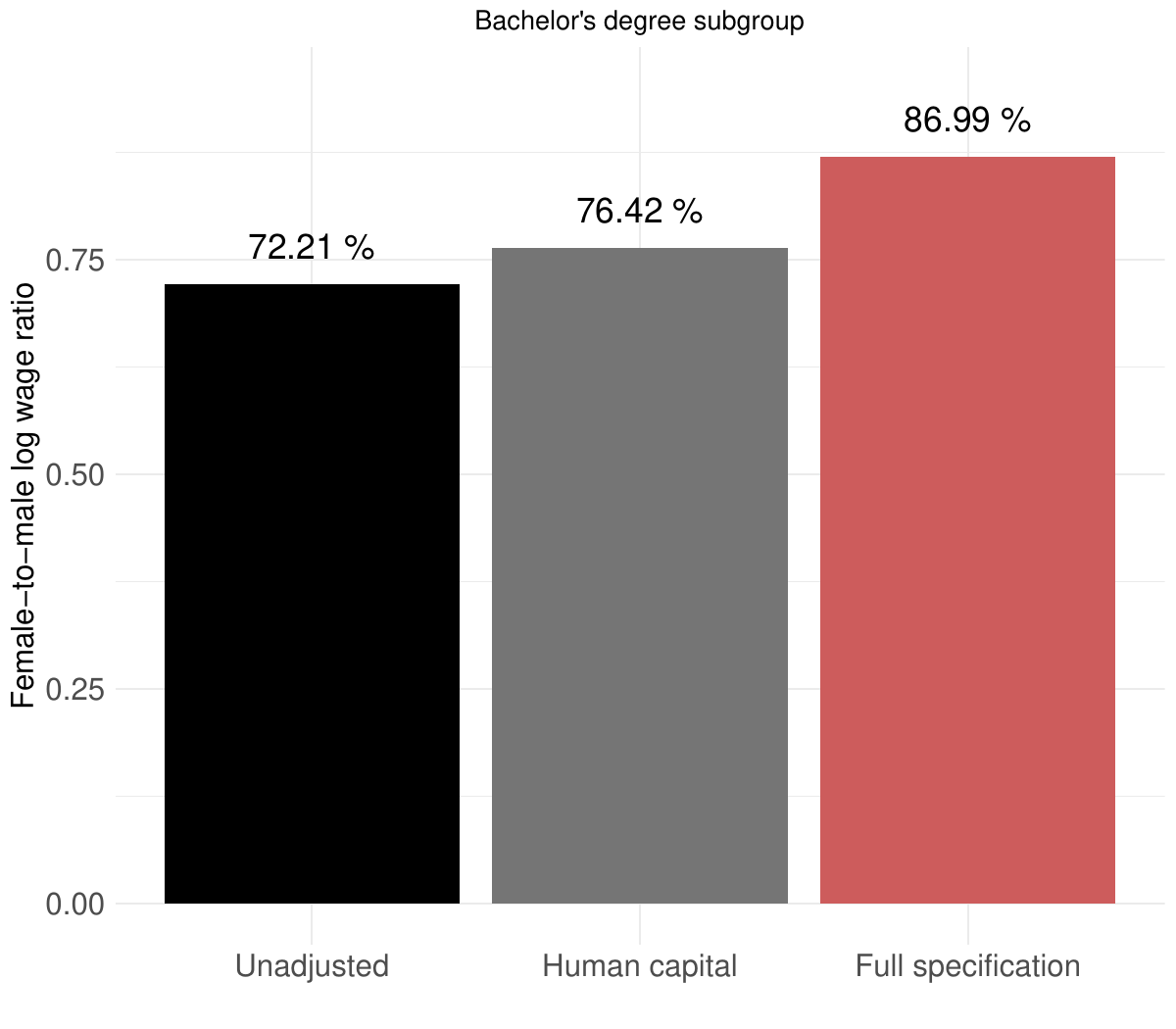}
  \end{minipage}
	\caption{Unconditional and conditional female-to-male mean (log) wage ratios.} 
	\label{fmratio}
		\begin{justify} 	\footnotesize
The barplots indicate the female-to-male log wage ratios, i.e., the quotient $\exp(\overline{x}_f \gamma_f)/\exp(\overline{x}_f \gamma_m)$, with regression coefficients $\gamma_f$ and $\gamma_m$ from a regression separately performed for the female and male observations. The unadjusted wage ratio corresponds to $\exp(\overline{\log(w_f)})/\exp(\overline{\log(w_m)})$. The human capital specification includes the regressors years of education, experience and experience squared. In the full regression specification, variables on industry, occupation, the number of hours worked each week and, for the bachelor's degree data only, the field of college major were included. In the regressions, we additionally controlled for English language ability; veteran status; U.S. census region; race; Hispanic origin; binary variables indicating if a person lived in the same household with a biological, adopted or stepchild children of age 4 or younger and, respectively, of age 18 or younger; metropolitan statistical area; and marital status. 	
	\end{justify}
	\end{figure}

We compared our results on a heterogeneous gender wage gap to those obtained from a traditional Oaxaca-Blinder decomposition and, in doing so, based the log wage ratio analysis on that in \cite{blau2016}, Chapter 2.1. The abovementioned decrease of the gender wage gap in terms of human capital characteristics in the high school degree subgroup were in concordance with the estimated female-to-male wage ratios shown in  Figure \ref{fmratio}.  Accordingly, the female-to-male wage ratio was slightly smaller if we conditioned on human capital factors. The resulting wage ratios were 77\% if we controlled for human capital characteristics and 78\% if we considered the unconditional gap. If we conditioned on additional individual characteristics including occupation, industry and hours worked, among others, average wages of female employees were around 17\% lower than wages of the male employees. The 17\% wage loss corresponds to the ``average female residual from the male wage equation'' in \cite{blau2016} (p. 800).  The patterns observed for the sample of academics revealed that conditioning on human capital factors lifted the female-to-male wage ratio from a level of 72\% to 76\%. Including additional individual characteristics led to an estimated wage ratio of 87\% corresponding to a residual wage gap of approximately 14\%. The residual wage gap estimates were slightly larger than in the study by \cite{blau2016} due to different data sources and different sample definitions.

\clearpage

\subsection{Model and methodology: A heterogeneous U.S. gender wage gap} \label{methodology}
We start with a basic log wage regression where the coefficient $\beta$ measures the relative difference in pay that emerges between men and women if one controls for the effects of observable characteristics. In the following we use a gender variable that is 1 if a person is female and $0$ if male.
\begin{align} \label{basicwage}
\ln{w_i} &= \alpha + \beta \cdot \text{gender}_i + x_i'\gamma + \varepsilon_i ,
\end{align}
Estimation of $\beta$ in Equation \ref{basicwage} results in an average gender wage gap that is of the same magnitude for all women irrespective of their observable characteristics. 
In order to model heterogeneity, we extend the basic wage equation in \ref{basicwage} and let the gender coefficient $\beta = \beta(x_i)$ be a function of individual  characteristics.  
\begin{align} \label{extendedwage}
\ln{w_i} &= \alpha + \beta(x_i) \cdot \text{gender}_i + z_i' \delta + \varepsilon_i.
\end{align}
The $\beta(x_i)$ coefficient can be a linear or a more complicated function of the $p_1$ observable characteristics $x_i$, for instance using transformations with splines or polynomials of higher order to approximate complex relationships of gender and the other explanatory variables. The covariates $z_i$ in  wage Equation \ref{extendedwage} are natural or constructed regressors. In our empirical application, we approximated $\beta(x_i)$ with a linear function of the regressors, i.e., $\beta(x_i) = \sum_{j=1}^{p_1} \beta_j  \cdot x_{j,i} $ and $z_i$ comprised all two-way interactions of the initial covariates $x_i$, including a constant. 
With this specification, the estimated model corresponded to 
\begin{align} \label{interactedwage}
\ln{w_i} &= \alpha +  \sum_{j=1}^{p_1} \left(\beta_j \cdot   x_{i,j}    \right) \cdot \text{gender}_i   +  z_i'\delta + \varepsilon_i .
\end{align}
We consider $p_1$ initial characteristics $x_i$ with corresponding coefficients $\beta_j$, $j = 1, \ldots, p_1$, that enter $\beta(x_i)$. Together with the dimension $p_2$ of $z_i$ and the corresponding vector of coefficients $\delta$, the overall number of parameters to be estimated is $p = p_1 + p_2 + 1$.  

A negative $\beta_j$, $\beta_j<0$, is interpreted as an increase of the absolute value of the wage gap. Hence, by default and in line with the public discussion, ``gender wage gap'' is interpreted as lower payment for women, although the opposite might be observed in the data.

\subsection{Methodological background: valid post-selection inference in high dimensions}  \label{inference}
Studying the heterogeneity of the gender wage gap in the presented model requires a rich set of observable characteristics and, hence, modern statistical methods to deal with high-dimensional data. 
We estimated the wage Equation \ref{extendedwage} with the lasso  and based inference on the double-selection approach of \cite{BelloniChernozhukovHansen2011}  that, in combination with the work by \cite{BCK-LAD} provides a uniformly valid inference framework for a vector of  ``target'' coefficients after model selection. We refer to the resulting estimators in the main text and the supplementary material  as ``double lasso''.
 Under a set of assumptions including sparsity, it is possible to perform valid post-selection inference even in cases where the number of regressors ($p$) exceeds the number of observations ($n$). In the heterogeneous wage gap regression, the lasso estimators $(\hat{\alpha}, \widehat{\beta}(x_i), \widehat{\delta})$ are defined as the solutions to
\begin{align}
 \label{lasso}
\left(\widehat{\alpha}, \widehat{\beta}(x_i), \widehat{\delta} \right)   \in & \\
 \arg \min_{\alpha, \beta(x_i), \delta}    \left\{ \frac{1}{n}\sum_{i=1}^{n}    \right. & \left. \left( \ln{w_i} - \left( \alpha + \beta(x_i) \cdot \text{\normalsize gender}_i + z_i'\delta   \right) \right)^2 + \right. \nonumber \\ 
  & \left.  \frac{\lambda}{n}  \parallel \Psi \left( \beta(x_i),  \delta' \right)'\parallel_1 \right\}  \nonumber ,%
\end{align}
with $\beta(x_i) = \sum_{j=1}^{p_1} \beta_j \cdot x_{j,i}$ and $\Psi$ a diagonal matrix with data-dependent weights.  
The lasso as initially developed in \cite{T1996} introduces a penalization by the $l_1$-norm of the coefficients to the least squares problem and serves as a variable selection device. Under the assumptions that sparsity in the data generating process holds,  solutions obtained with the lasso are sparse, i.e., only relatively few, say $s$, of the $p$ candidate regressors have explanatory power for the outcome variable.  Sparsity avoids overfitting that is likely to arise in ordinary least squares regression with many regressors.
Estimating \ref{lasso} requires a choice of the penalty $\lambda$. Frequently, $\lambda$ is set by (k-fold) cross-validation. However, since cross-validation is not backed by theoretical arguments in a high-dimensional setting and computationally expensive,  we determined $\lambda$ by the theory-based rule of \cite{BCCH12} which is also applicable to the case of heteroscedasticity. An  intuitive introduction to the lasso and the reasoning of the penalty choice can be found in \cite{hdsm2011}.
To be more exact, we estimated a post-selection version of the lasso, the so-called ``post-lasso'' \citep{BC-PostLASSO}, i.e., we re-estimated the coefficients that have been selected by the lasso  with  ordinary least squares regression. The penalization imposed by the lasso causes a shrinkage of the coefficients towards zero. A part of this bias can be alleviated by using post-lasso. 

Post-selection inference, i.e., inference on coefficients after a model selection stage, has been an active  research area in the statistics literature in the last years. In general, simply conducting ordinary least squares inference after the lasso selection step, i.e., perform inference as if there was no selection, does not result in a valid inference procedure unless perfect model selection is achieved. However, the latter is only guaranteed under strict assumptions, for instance a so-called \textit{beta-min} assumption imposing that the non-zero coefficients are well-distinguishable from zero. 

The challenge for valid inference under a model selection step with the lasso or other ``machine learning'' methods is to avoid model selection mistakes for variables that are both correlated with the outcome and the target variables of interest (i.e., incorrectly excluding ``confounders'' from the model). The failure of inference validity due to that omitted variable bias is illustrated in an intuitive example in \cite{BelloniChernozhukovHansen2011}. 
The double-selection approach of \cite{BelloniChernozhukovHansen2011}, however, and more generally, estimation based on orthogonalized moment conditions as in \cite{BCK-LAD}, offer an opportunity to resolve the problems of inference after a model selection stage. The idea of the method is to introduce an auxiliary lasso regression for every target coefficient to ensure that only moderate selection mistakes might occur.  The idea of the method is to introduce an auxiliary lasso regression for every target coefficient to ensure that only moderate selection mistakes might occur.  Double-selection proceeds as follows: 
\begin{enumerate}
\item[1.] For each of the $p_1$ target variables, estimate a lasso regression of the dependent variable in Equation \ref{extendedwage}, $\ln w_i$, on regressors $z_i$ and the remaining targets. The target variables correspond to the interactions with gender in Equation \ref{extendedwage}.
\item[2.] Estimate an auxiliary lasso regression of each of the $p_1$ target variables on all remaining independent variables as regressors. 
\item[3.] Equation \ref{extendedwage} is re-estimated with ordinary least squares regression with all variables being included that have been selected in either the first or the auxiliary regression steps.
\end{enumerate}
 For more details on the double-selection approach and post-selection inference on a set of target coefficients, we refer to \cite{BelloniChernozhukovHansen2011} and \cite{BCK-LAD}. 
  Following \cite{BelloniChernozhukovHansen2011} 
	and using  asymptotic normality of the double-selection estimators, it is possible to show that under sparsity, $\beta(x_i)$ as estimated by double lasso  asymptotically follows  a normal distribution 
\begin{align} \label{k}
\sqrt{n} \left( \hat \beta(x_i) - \beta_0 (x_i)\right) \leadsto^d N \left(0, x_i' \Omega x_i \right),
\end{align}
with variance-covariance matrix $\Omega$ of the $\hat{\beta}_j$ in $\hat{\beta}(x_i)$ which can be estimated according to \cite{BelloniChernozhukovHansen2011}. 

As the number of target parameter in the heterogeneous gender wage model is large, it is necessary to adjust for multiple testing. 
We implement the multiplier bootstrap procedure developed in \cite{CCK12} and \cite{CCK:AOS13} to construct uniformly valid confidence intervals for $\beta(x_i)$ and perform a valid joint test for the marginal effect targets $\beta_j$, $j=1,...,p_1$, as suggested in \cite{BCK-LAD}. Moreover, to adjust $p$-values in the joint hypothesis test, we apply the stepdown procedure of \cite{romanowolf2005, romanowolf2, romanowolf2016}, as recently established in \cite{CCK:AOS13} and \cite{BCK-LAD}. For a more detailed presentation of the Romano-Wolf stepdown procedure and the underlying algorithm to construct $p$-values, we refer to \cite{siminf}. 

Tables \ref{fullresultsHS1} to \ref{fullresultsBA4} present all double lasso estimates irrespective of their significance. $p$-values were obtained from a joint significance tests of all $\beta_j$ coefficients in $\beta(x_i)$  from Equation \ref{extendedwage} using the multiplier bootstrap procedure suggested in \cite{BCK-LAD} with 1000 repetitions in combination with the stepdown procedure of \cite{romanowolf2005, romanowolf2, romanowolf2016}. Moreover, Table \ref{resultsOCC} presents ordered estimates of occupational effects separately for both educational subgroups to illustrate wage gap heterogeneity in occupation. The R-code of the entire analysis will be provided guaranteeing full replicability of all results.

\subsection{Empirical methods in the context of the gender wage gap} \label{literature}

Traditionally, decomposition methods as initially introduced in \cite{oaxaca1973} and \cite{blinder1973} are used to assess the gender wage gap empirically. A detailed and comprehensive overview on decomposition methods and recent extensions thereof is provided in \cite{fortin2011decomposition}. 
The objective of the Oaxaca-Blinder decomposition is to distinguish whether the overall wage difference between men and women emerges due to gender differences in observable characteristics or due to a different valuation of these characteristics in the labor market, sometimes referred to as a ``wage structure effect'' \citep{fortin2011decomposition}. An example for the first effect is a situation with higher labor market experience, on average, for men than for women. The second effect emerges from the difference of the regression coefficients from two wage regressions that are separately estimated for male and female observations in the data. An example for the structural effect is a situation with higher returns to labor market experience for men than for women. A gender difference in valuations of labor market characteristics is often considered as an indicator of discrimination, although it might also reflect non-discriminatory effects, for instance unobserved productivity effects \citep{blau2016}. 
Recently, the econometric literature has developed innovative methodological extensions of the basic Oaxaca-Blinder decomposition  that base upon quantile regression, for instance \cite{chernozhukov2013counter}. These methods are able to detect heterogeneous patterns of the gender gap at different points of the income distribution. For instance, the gender wage gap was found to be more pronounced at the top of the income distribution than in the middle or at the bottom \cite{blau2016}.

\cite{goldin2014} provides a recent study of the gender wagegap that can be related to our approach. In the empirical analysis of \cite{goldin2014} that is based on ACS data, an ordinary least squares regression of an extended wage equation is estimated  that included interactions of gender with a large number (i.e., 469)  of occupation dummies. Being based on a theoretical argument, the gender wage gap is allowed to vary across occupational categories, and, hence, the focus of the  heterogeneity analysis is on variation by occupation. The results of \cite{goldin2014} illustrate the variation of the gender wage gap in an appealing way. Unfortunately, the significance of the effects is not reported. Under statistical considerations, however, the question of joint significance of heterogeneous effects is of great importance: If the number of tested hypotheses is large, adjustments for simultaneously testing multiple hypotheses are required to provide valid tests, confidence bands and $p$-values. 

An approach that is related to our econometric framework has been recently developed by \cite{sortedeffects}. Similar to the quantile plots, which we present in the main text, the so-called sorted effects methods provides estimates and confidence bands for an ordered sequence of partial effects that quantify heterogeneity in terms of observational characteristics. Indeed, the quantile plots in the main text coincide with the sorted effects if ordinary least squares regression is employed and an appropriate structural regression model is chosen in both approaches. Whereas the interpretation of our quantile plots and the sorted effects is similar, the approach to analyze heterogeneity as a variation of the partial effects in terms of observed variables in \cite{sortedeffects} differs from our analysis. The  so-called classification analysis in \cite{sortedeffects} provides an inferential framework for testing differences in observational characteristics of individuals in the most and least affected subgroup. In contrast, the focus in our study is on the variation of the gender wage gap estimate according to observational characteristics, in other words variation of $\beta(x)$ according to differences in $x$. Moreover, we base estimation of the regression equation on the lasso estimator and the double-selection framework of \cite{BelloniChernozhukovHansen2011}.



\subsection{Literature review: the gender wage gap and recent developments}

A great number of empirical studies have focused on the gender wage gap, its determinants and its development over time and the life cycle. Due to the richness of the gender gap literature, we restrict attention to the literature on the gender gap in earnings and its determinants. \cite{blau2016} provide an extensive and detailed review of various explanations of the gender wage gap together with an empirical reassessment of many theories. 

The second half of the 20th century was characterized by a substantial convergence of the gender wage gap paralleled by a considerable convergence of men and women in terms of education, labor market experience and participation,  and occupational choices, among others \citep{goldin2014, blau2016}.
A large part of the reduction of the gender wage gap that began in the 1980s and still continues until today, although in a less steady and slower manner, is attributed to the convergence in traditional human capital factors. Today, women achieve higher levels of education than men and almost the same levels of actual experience, on average. 
In a recent analysis, \cite{blau2016} provide evidence that gender differences in observable characteristics such as experience, occupation and industry variables,  explained two thirds of the total gender gap in 2010. 
%
As gender differences in terms of traditional human capital characteristics have diminished over time, these factors have become less important in explaining the gender wage gap. For instance, in the decomposition of \cite[Table 4]{blau2016}, differences in human capital characteristics could only explain 13\% of the total gender wage gap in 2010 compared to 25\% in 1980.

Alternative explanations have been developed in the labor economics literature. A recently proposed reasoning by \cite{goldin2014}  focuses on the structure of jobs. Temporal flexibility, referring to factors like the total number of hours worked and the time when they are provided, translates into a convex relationship of working hours and the salary. Since women typically value flexibility more than men because of a greater involvement in child rearing, gender inequality in earnings is expected to be more pronounced  in inflexible occupations. 
\cite{goldin2014} presents evidence that the wage gap was larger and increased over the lifecycle in inflexible occupations, for example in the area of business or law, compared to more flexible occupations like pharmacy, science or technology. Moreover, in less flexible occupations, the gender wage gap was found to increase with the number of hours worked due to a more convex hours-earnings relationship. The explanatory power of occupations  for the gender gap, together with industries, was also empirically confirmed in the analysis of \cite{blau2016}. 

The argumentation of \cite{goldin2014} and other studies is related to the fact that women are more likely to interrupt their work life because of having children and a greater responsibility in child rearing. Using data on actual labor market experience, 
\cite{blau2016} emphasize the role of work life interruptions for the wage gap. In general, the effect of interruptions is relatively difficult to assess in empirical studies due to limitated availability of actual labor market experience in many data sets.  Moreover, explanations in favor of a ``family'' or ``motherhood penalty'' \citep{waldfogel1998, siglerushton2007} have been proposed and confirmed empirically implying that mothers tend to experience larger wage gaps than women without children.  Recent studies, which mainly use administrative data from Scandinavian countries, assess the dynamics of the motherhood penalty over women's working history \citep{kleven2018, angelov2016, albrecht2018}. A recently pubilshed study by \cite{butikofer2018} focuses on the motherhood penalty in high-paying jobs in Norway and assesses differences across four  occupational categories. The analysis is based on the flexibility argumentation of \cite{goldin2014} and finds an association of greater motherhood penalties and occupations with  lower flexibility. 

Furthermore, behavioral explanations suggest that psychological attributes and norms, for example weak preferences for competition and negotiations, cause gender differences in wages \citep{mueller2006, manning2008}. However, \cite{blau2016} conclude that these explanations cannot explain a large fraction of the gender wage gap and that further empirical non-laboratory evidence with stronger external validity is required to assess the importance of these theories. 

Finally, taste-based or statistical discrimination is a potential source of the gender wage gap. The \textit{unexplained gender wage gap} from an Oaxaca-Blinder decomposition is frequently taken as a measure of discrimination. However, the unexplained gap might as well be the result of  unobserved factors related to productivity. Hence, there is no unambiguous  empirical evidence  of discrimination that is based on observational data. Real-world  experiments point at a discrimination against women and mothers, for example \cite{neumark1996, correll2007}. \cite{blau2016} conclude that a part of the convergence of the wage gap in the 20th century might be explained by reduced discrimination against women in the labor market.

\subsection{Relation to Oaxaca-Blinder Decomposition}

\begin{remark}[Relation to the Oaxaca-Blinder Decomposition] \label{oaxaca}
 In  the case of a linear function $\beta(\cdot)$ and the covariate vector $z_i$ comprising all two-way interactions of the initial covariates $x_i$, the heterogeneous gender gap model can be related to the Oaxaca-Blinder decomposition. Suppose, one estimates the wage regression 
\begin{align*}
\ln{w_i} &= \alpha + \beta(z_i) \cdot \text{gender}_i + z_i' \delta + \varepsilon_i.
\end{align*}
Then, the mean of $\beta(z_i)$ corresponds to the negative of the total unexplained gender gap (``the structural effect'') from an Oaxaca-Blinder decomposition, in other words the part of the gender wage gap that emerges due to different valuations of labor market characteristics for men and women
\begin{align*}
\overline{\beta(z_i)} &= \frac{1}{n_f} \sum_{i=1}^{n_f} \beta(z_i) = - \overline{z}_f' \left(\gamma_m - \gamma_f \right),
\end{align*}
with $(\gamma_m, \gamma_f)$ being the coefficients obtained from the regressions performed separately for the subset of men ($m$) and women ($f$), $n_f=\sum_i^{n} \text{gender}_i$ being the number of female observations, and $\overline{z}_f$ being the matrix collecting the mean values of the interacted initial observable characteristics of women,  $x_i$.
\end{remark}

\subsection{Interpretation of $\beta(x_i)$} \label{interpret}

The proposed  model captures the heterogeneity of the gender gap in the function $\beta(x_i)$. 
We interpret a negative $\beta(x_i)$ as the approximate gender wage gap experienced by a woman with characteristics $x_i$ on average. Hence, a woman in the subgroup of individuals with characteristics $x_i$ earns approximately $\beta(x_i) \cdot 100\%$ less than a male employee in the same subgroup, in other words a man with the same educational attainment, working in the same industry and occupation, and so on.

We are not only interested in estimating the gender pay gap for every woman in the sample, but also in assessing the determinants of the wage gap. In a linear specification including a constant, $\beta(x_i)= x_i'\beta$,  the $j$th component of $\beta$, $\beta_j$, indicates the marginal change of the wage gap for a woman differing only with regard to this variable.   
In case a regressor is continuous, the gender gap change due to a marginal change in variable $x_j$ is ceteris paribus
\begin{align} \label{betas2}
\frac{\partial \beta(x_i)}{\partial x_j} &= \beta_j .
\end{align}

\clearpage 



\include{tables/resulttables_lasso/resulttable_OCC}

\include{tables/resulttables_lasso/resulttableHSfull}

\include{tables/resulttables_lasso/resulttableBAfull}

\clearpage
\footnotesize
\pagebreak
\bibliographystyle{plainnat}
\bibliography{mybib}

%% file: tables/resulttables_lasso/resulttable_OCC.tex
\begin{table}
\centering
        \begin{tabular}{lrr   l rr}
 \hline \\[-0.8ex]
  \multicolumn{3}{c}{High school degree subgroup} &\multicolumn{3}{c}{Bachelor's degree subgroup}  \\[0.8ex]
Variable & Coefficient & $p$-value  & Variable & Coefficient & $p$-value \\ 
 \hline  \\[-0.8ex]
Educ/Training/Libr & -0.1836 & 0.0000 & \textcolor[rgb]{0.4,0.4,0.4}{Healthc Supp} &\textcolor[rgb]{0.4,0.4,0.4}{-0.1022} &\textcolor[rgb]{0.4,0.4,0.4}{0.3950} \\ 
 \textcolor[rgb]{0.4,0.4,0.4}{Extract} &\textcolor[rgb]{0.4,0.4,0.4}{-0.1448} &\textcolor[rgb]{0.4,0.4,0.4}{0.9600} &  \textcolor[rgb]{0.4,0.4,0.4}{Milit Specific} &\textcolor[rgb]{0.4,0.4,0.4}{-0.0799}&\textcolor[rgb]{0.4,0.4,0.4}{0.9930} \\ 
  Prod & -0.0974 & 0.0000 & \textcolor[rgb]{0.4,0.4,0.4}{Farm/Fish/Forestry} &\textcolor[rgb]{0.4,0.4,0.4}{-0.0498} &\textcolor[rgb]{0.4,0.4,0.4}{1.0000}\\ 
 \textcolor[rgb]{0.4,0.4,0.4}{Arts/Design/Entert/}&\textcolor[rgb]{0.4,0.4,0.4}{-0.0304} &\textcolor[rgb]{0.4,0.4,0.4}{0.9200} &   Office/Administr Supp & -0.0465 & 0.0000 \\ 
	\textcolor[rgb]{0.4,0.4,0.4}{Sports/Media}  & & & Healthc Pract/Technic & -0.0407 & 0.0260 \\
 \textcolor[rgb]{0.4,0.4,0.4}{Sales} &\textcolor[rgb]{0.4,0.4,0.4}{-0.0187} &\textcolor[rgb]{0.4,0.4,0.4}{0.7210}  &  \textcolor[rgb]{0.4,0.4,0.4}{Financ Spec} &\textcolor[rgb]{0.4,0.4,0.4}{-0.0348} &\textcolor[rgb]{0.4,0.4,0.4}{0.0690}  \\ 
 \textcolor[rgb]{0.4,0.4,0.4}{Financ Spec} &\textcolor[rgb]{0.4,0.4,0.4}{-0.0127} &\textcolor[rgb]{0.4,0.4,0.4}{0.9760} &  \textcolor[rgb]{0.4,0.4,0.4}{Pers Care/Serv} &\textcolor[rgb]{0.4,0.4,0.4}{-0.0287} &\textcolor[rgb]{0.4,0.4,0.4}{1.0000} \\ 
	\textcolor[rgb]{0.4,0.4,0.4}{Build/Grounds Clean/} &\textcolor[rgb]{0.4,0.4,0.4}{-0.0108} &\textcolor[rgb]{0.4,0.4,0.4}{0.9710}  & \textcolor[rgb]{0.4,0.4,0.4}{Build/Grounds Clean/}&\textcolor[rgb]{0.4,0.4,0.4}{-0.0248} &\textcolor[rgb]{0.4,0.4,0.4}{1.0000}  \\
  \textcolor[rgb]{0.4,0.4,0.4}{Mainten}  &  & &  \textcolor[rgb]{0.4,0.4,0.4}{Mainten} & &   \\ 
 \textcolor[rgb]{0.4,0.4,0.4}{Transp} &\textcolor[rgb]{0.4,0.4,0.4}{-0.0085} &\textcolor[rgb]{0.4,0.4,0.4}{0.9760} &  \textcolor[rgb]{0.4,0.4,0.4}{Sales}&\textcolor[rgb]{0.4,0.4,0.4}{-0.0162} &\textcolor[rgb]{0.4,0.4,0.4}{0.9980}  \\ 
& & &  \textcolor[rgb]{0.4,0.4,0.4}{Food Prepar/Serving} &\textcolor[rgb]{0.4,0.4,0.4}{-0.0011} &\textcolor[rgb]{0.4,0.4,0.4}{1.0000} \\

& & & & & \\
 \textcolor[rgb]{0.4,0.4,0.4}{Archit/Engin} &\textcolor[rgb]{0.4,0.4,0.4}{0.0189} &\textcolor[rgb]{0.4,0.4,0.4}{0.9760}& \textcolor[rgb]{0.4,0.4,0.4}{Prod} &\textcolor[rgb]{0.4,0.4,0.4}{0.0065} &\textcolor[rgb]{0.4,0.4,0.4}{1.0000} \\ 
  \textcolor[rgb]{0.4,0.4,0.4}{Pers Care/Serv}&\textcolor[rgb]{0.4,0.4,0.4}{0.0200} &\textcolor[rgb]{0.4,0.4,0.4}{0.9200} &  \textcolor[rgb]{0.4,0.4,0.4}{Transp} &\textcolor[rgb]{0.4,0.4,0.4}{0.0228} &\textcolor[rgb]{0.4,0.4,0.4}{1.0000}  \\ 
 \textcolor[rgb]{0.4,0.4,0.4}{Comput/Math} &\textcolor[rgb]{0.4,0.4,0.4}{0.0246} &\textcolor[rgb]{0.4,0.4,0.4}{ 0.7830} &   Comput/Math & 0.0372 & 0.0020 \\ 
 \textcolor[rgb]{0.4,0.4,0.4}{Farm/Fish/Forestry} &\textcolor[rgb]{0.4,0.4,0.4}{0.0286} &\textcolor[rgb]{0.4,0.4,0.4}{ 0.9760} &   Bus Operat Spec & 0.0377 & 0.0110\\ 
 \textcolor[rgb]{0.4,0.4,0.4}{Food Prepar/Serving} &\textcolor[rgb]{0.4,0.4,0.4}{0.0290}&\textcolor[rgb]{0.4,0.4,0.4}{0.2360} &   Arts/Design/Entert/ & 0.0469 & 0.0330 \\ 
 \textcolor[rgb]{0.4,0.4,0.4}{Milit Specific} &\textcolor[rgb]{0.4,0.4,0.4}{0.0391} &\textcolor[rgb]{0.4,0.4,0.4}{0.9760} & Sports/Media & & \\ 
 \textcolor[rgb]{0.4,0.4,0.4}{Technic} &\textcolor[rgb]{0.4,0.4,0.4}{0.0419} &\textcolor[rgb]{0.4,0.4,0.4}{0.9200} &  \textcolor[rgb]{0.4,0.4,0.4}{Legal} &\textcolor[rgb]{0.4,0.4,0.4}{0.0495} &\textcolor[rgb]{0.4,0.4,0.4}{0.0810}  \\ 
  \textcolor[rgb]{0.4,0.4,0.4}{Protect Serv} & \textcolor[rgb]{0.4,0.4,0.4}{0.0479} &  \textcolor[rgb]{0.4,0.4,0.4}{0.0510} & Educ/Training/Libr & 0.0606 & 0.0000 \\ 
  Healthc Supp & 0.0530 & 0.0420 & Archit/Engin & 0.0620 & 0.0000 \\ 
  Bus Operat Spec & 0.0571 & 0.0030 &  Protect Serv & 0.0666 & 0.0110 \\ 
  Office/Administr Supp & 0.0635 & 0.0000 &  Life/Physical/Soc Sci. & 0.0719 & 0.0000 \\ 
 \textcolor[rgb]{0.4,0.4,0.4}{Life/Physical/Soc Sci.} &\textcolor[rgb]{0.4,0.4,0.4}{0.0703} &\textcolor[rgb]{0.4,0.4,0.4}{0.5360} &  \textcolor[rgb]{0.4,0.4,0.4}{Technic} &\textcolor[rgb]{0.4,0.4,0.4}{0.1126} &\textcolor[rgb]{0.4,0.4,0.4}{0.5270} \\ 
  Install/Mainten/Rep & 0.0742 & 0.0070 & \textcolor[rgb]{0.4,0.4,0.4}{Constr} &\textcolor[rgb]{0.4,0.4,0.4}{0.1469} &\textcolor[rgb]{0.4,0.4,0.4}{0.0810} \\ 
 Constr & 0.0895 & 0.0150    & Install/Mainten/Rep & 0.1496 & 0.0020  \\ 
   \textcolor[rgb]{0.4,0.4,0.4}{Legal} &\textcolor[rgb]{0.4,0.4,0.4}{0.1042} &\textcolor[rgb]{0.4,0.4,0.4}{0.3400} & Comm/Soc Serv &  0.1702 & 0.0000 \\ 
 Healthc Pract/Technic & 0.1075 & 0.0000 & & &  \\ 
  Comm/Soc Serv & 0.1205 & 0.0000 & & & \\ 
  \hline \\[0.8ex]
\end{tabular}
   \caption{Ordered occupational effects, high school degree and bachelor's degree data, DOUBLE LASSO.}
   \label{resultsOCC}
\begin{justify} \footnotesize
The table presents occupational effects for the high school degree subgroup (left) and the bachelor's degree  subgroup (right) in increasing order to provide a comparison of the occupational patterns observed for both subgroups. $p$-values are obtained from a joint test of all $\beta_j$ coefficients in $\beta(x_i)$ from Equation \ref{extendedwage} using the multiplier bootstrap procedure suggested in \cite{BCK-LAD} with 1000 repetitions in combination with the stepdown procedure of \cite{romanowolf2005}. 
Significant (printed black) and non-significant (printed gray) coefficients at a 5\% significance levels are presented. 
\end{justify}

\end{table}

%% file: tables/resulttables_lasso/resulttableHSfull.tex
\begin{table}[!thb]
\centering
\begin{tabular}{lrr}
  \hline \\[-0.8ex]
Variable & Estimate &  $p$-value  \\ [0.8ex]
  \hline \\ [-0.8ex]
constant & -0.0463 & 0.9070 \\ 
[0.8ex]  \multicolumn{3}{l}{\textit{Marital status}} \\ [0.8ex]
 Married, spouse present & -0.1096 & 0.0000 \\ 
  Married, spouse absent & -0.0737 & 0.0010 \\ 
  Separated & -0.0575 & 0.0030 \\ 
  Divorced & -0.0571 & 0.0000 \\ 
  Widowed & -0.0536 & 0.0700 \\ 
	[0.8ex] \multicolumn{3}{l}{\textit{English language ability}} \\[0.8ex]
Does not speak English & 0.0550 & 0.1600 \\ 
  Yes, speaks very well & 0.0111 & 0.9200 \\ 
  Yes, speaks well & 0.0172 & 0.8850 \\ 
  Yes, but not well & 0.0303 & 0.3400 \\ 
	[0.8ex] \multicolumn{3}{l}{\textit{ Race, ethnicity}} \\[0.8ex]
 Black/African American/Negro & 0.0789 & 0.0000 \\ 
  Chinese & 0.0819 & 0.0100 \\ 
  Other Asian or Pacific Islander & 0.0716 & 0.0000 \\ 
    Hispanic & 0.0115 & 0.9200 \\ 
	[0.8ex] \multicolumn{3}{l}{\textit{Veteran status}} \\[0.8ex]
Veteran & 0.0429 & 0.0140 \\ 
	[0.8ex] \multicolumn{3}{l}{\textit{Industry}} \\[0.8ex]
AGRI & -0.0419 & 0.8540 \\ 
  MINING & -0.0656 & 0.8540 \\ 
  CONSTR & -0.0511 & 0.1330 \\ 
  MANUF & -0.0283 & 0.4020 \\ 
  TRANS & -0.0535 & 0.0030 \\ 
  RETAIL & -0.0444 & 0.0150 \\ 
  FINANCE & -0.0493 & 0.0180 \\ 
  BUISREPSERV & -0.0433 & 0.0640 \\ 
  PERSON & -0.0384 & 0.3860 \\ 
  ENTER & -0.0281 & 0.9200 \\ 
  PROFE & -0.0742 & 0.0000 \\ 
  ADMIN & -0.0527 & 0.0140 \\ 
  MILIT & 0.1145 & 0.2650 \\  \hline
\end{tabular}
\caption{Full DOUBLE LASSO results (1/3), high school degree data.}
\label{fullresultsHS1}
\end{table}

\clearpage

\begin{table}[thb]
\centering
\begin{tabular}{lrr}
  \hline \\[-0.8ex]
Variable & Estimate &  $p$-value  \\ [0.8ex]
  \hline \\ 
[0.8ex] \multicolumn{3}{l}{\textit{Occupation}} \\[0.8ex] 
Bus Operat Spec & 0.0571 & 0.0030 \\ 
  Financ Spec & -0.0127 & 0.9760 \\ 
  Comput/Math & 0.0246 & 0.7830 \\ 
  Archit/Engin & 0.0189 & 0.9760 \\ 
  Technic & 0.0419 & 0.9200 \\ 
  Life/Physical/Soc Sci. & 0.0703 & 0.5360 \\ 
  Comm/Soc Serv & 0.1205 & 0.0000 \\ 
  Legal & 0.1042 & 0.3400 \\ 
  Educ/Training/Libr & -0.1836 & 0.0000 \\ 
  Arts/Design/Entert/Sports/Media & -0.0304 & 0.9200 \\ 
  Healthc Pract/Technic & 0.1075 & 0.0000 \\ 
  Healthc Supp & 0.0530 & 0.0420 \\ 
  Protect Serv & 0.0479 & 0.0510 \\ 
  Food Prepar/Serving & 0.0290 & 0.2360 \\ 
  Build/Grounds Clean/Mainten & -0.0108 & 0.9710 \\ 
  Pers Care/Serv & 0.0200 & 0.9200 \\ 
  Sales & -0.0187 & 0.7210 \\ 
  Office/Administr Supp & 0.0635 & 0.0000 \\ 
  Farm/Fish/Forestry & 0.0286 & 0.9760 \\ 
  Constr & 0.0895 & 0.0150 \\ 
  Extract & -0.1448 & 0.9600 \\ 
  Install/Mainten/Rep & 0.0742 & 0.0070 \\ 
  Prod & -0.0974 & 0.0000 \\ 
  Transp & -0.0085 & 0.9760 \\ 
  Milit Specific & 0.0391 & 0.9760 \\ 
		[0.8ex] \multicolumn{3}{l}{\textit{U.S. Census region}} \\[0.8ex]
  Middle Atlantic Division & -0.0110 & 0.9580 \\ 
  East North Central Div. & -0.0086 & 0.9760 \\ 
  West North Central Div. & -0.0065 & 0.9760 \\ 
  South Atlantic Division & 0.0016 & 0.9820 \\ 
  East South Central Div. & -0.0254 & 0.5560 \\ 
  West South Central Div. & -0.0311 & 0.1070 \\ 
  Mountain Division & -0.0010 & 0.9820 \\ 
  Pacific Division & 0.0200 & 0.6990 \\ 
		\hline
\end{tabular}
\caption{Full DOUBLE LASSO results (2/3), high school degree data.}
\label{fullresultsHS2}
\end{table}

\clearpage

\begin{table}[thb]
\begin{tabular}{lrr}
  \hline \\[-0.8ex]
Variable & Estimate &  $p$-value  \\ [0.8ex]
  \hline \\ 
[0.8ex] \multicolumn{3}{l}{\textit{Metropolitan statistical area}} \\[0.8ex]
	msa & 0.0139 & 0.4120 \\ 
		[0.8ex] \multicolumn{3}{l}{\textit{Child}} \\[0.8ex]
  Age $18$ or younger & -0.0507 & 0.0000 \\ 
  Age $4$ or younger & 0.0289 & 0.0180 \\ 
  [0.8ex] \multicolumn{3}{l}{\textit{Usual hours worked per week}} \\[0.8ex] 
  40 to 49 & -0.0456 & 0.0000 \\ 
  50 to 59 & -0.0374 & 0.0150 \\ 
  60 to 69 & -0.0534 & 0.0150 \\ 
  $>$ 70 & -0.1186 & 0.0000 \\ 
	  [0.8ex] \multicolumn{3}{l}{\textit{Years of education}} \\[0.8ex]
  yos & -0.0026 & 0.9200 \\ 
	  [0.8ex] \multicolumn{3}{l}{\textit{Experience}} \\[0.8ex]
  exp & -0.0040 & 0.0010 \\ 
  exp2 & 0.0000 & 0.3180 \\ 
			\hline
\end{tabular}
\caption{Full DOUBLE LASSO results (3/3), high school degree data.}
\label{fullresultsHS3}
\begin{justify}
 Tables \ref{fullresultsHS1} to \ref{fullresultsHS3} present complete results from post-lasso estimation using double selection (DOUBLE LASSO) obtained for the high school degree subsample. $p$-values are obtained from a joint test of all $\beta_j$ coefficients in $\beta(x_i)$ from Equation \ref{extendedwage} using the multiplier bootstrap procedure suggested in \cite{BCK-LAD} with 1000 repetitions in combination with the stepdown procedure of \cite{romanowolf2005}. 
\end{justify}
\end{table}

%% file: tables/resulttables_lasso/resulttableBAfull.tex
\begin{table}[thb]
\centering
\begin{tabular}{lrr}
  \hline \\[-0.8ex]
Variable & Estimate  & $p$-value \\ [0.8ex]
  \hline \\ [-0.8ex]
constant &  0.0428 & 1.0000 \\ 
 [0.8ex]  \multicolumn{3}{l}{\textit{Marital status}} \\ [0.8ex]
Married, spouse present & -0.0973 & 0.0000 \\ 
  Married, spouse absent & -0.0535 & 0.2630 \\ 
  Separated & -0.1205 & 0.0000 \\ 
  Divorced & -0.0548 & 0.0000 \\ 
  Widowed & -0.1152 & 0.0110 \\ 
	[0.8ex]  \multicolumn{3}{l}{\textit{English language ability}} \\ [0.8ex]
Does not speak English & 0.0221 & 1.0000 \\ 
  Yes, speaks very well & -0.0022 & 1.0000 \\ 
  Yes, speaks well & 0.0392 & 0.3720 \\ 
  Yes, but not well & 0.0030 & 1.0000 \\ 
	[0.8ex]  \multicolumn{3}{l}{\textit{Race, ethnicity}} \\ [0.8ex]
Black/African American/Negro & 0.0679 & 0.0000 \\ 
  Chinese & 0.0589 & 0.0020 \\ 
  Other Asian or Pacific Islander & 0.0437 & 0.0010 \\ 
  Hispanic & 0.0070 & 1.0000 \\ 
[0.8ex]  \multicolumn{3}{l}{\textit{Veteran status}} \\ [0.8ex]
Veteran & 0.0204 & 0.9930 \\ 
	[0.8ex]  \multicolumn{3}{l}{\textit{Industry}} \\ [0.8ex]
AGRI & -0.0655 & 0.8910 \\ 
  MINING & -0.0672 & 0.9680 \\ 
  CONSTR & -0.0310 & 0.9990 \\ 
  MANUF & -0.0040 & 1.0000 \\ 
  TRANS & 0.0217 & 1.0000 \\ 
  RETAIL & -0.0216 & 1.0000 \\ 
  FINANCE & -0.0799 & 0.0000 \\ 
  BUISREPSERV & -0.0557 & 0.0450 \\ 
  PERSON & -0.0564 & 0.7250 \\ 
  ENTER & -0.0630 & 0.6300 \\ 
  PROFE & -0.0668 & 0.0010 \\ 
  ADMIN & -0.0091 & 1.0000 \\ 
  MILIT & 0.1167 & 0.2040 \\ 
		\hline	
	\end{tabular}
\caption{DOUBLE LASSO results (1/4), bachelor's degree data.}
\label{fullresultsBA1} 
\end{table}

\clearpage

\begin{table}[thb]
	\vspace{0.1cm}
\centering
\begin{tabular}{lrr}
  \hline \\[-0.8ex]
Variable & Estimate & $p$-value  \\ [0.8ex]
  \hline \\ 
[0.8ex]  \multicolumn{3}{l}{\textit{Occupation}} \\ [0.8ex]
  Bus Operat Spec & 0.0377 & 0.0110 \\ 
  Financ Spec & -0.0348 & 0.0690 \\ 
  Comput/Math & 0.0372 & 0.0020 \\ 
  Archit/Engin & 0.0620 & 0.0000 \\ 
  Technic & 0.1126 & 0.5270 \\ 
  Life/Physical/Soc Sci. & 0.0719 & 0.0000 \\ 
  Comm/Soc Serv & 0.1702 & 0.0000 \\ 
  Legal & 0.0495 & 0.0810 \\ 
  Educ/Training/Libr & 0.0606 & 0.0000 \\ 
  Arts/Design/Entert/Sports/Media & 0.0469 & 0.0330 \\ 
  Healthc Pract/Technic & -0.0407 & 0.0260 \\ 
  Healthc Supp & -0.1022 & 0.3950 \\ 
  Protect Serv & 0.0666 & 0.0110 \\ 
  Food Prepar/Serving & -0.0011 & 1.0000 \\ 
  Build/Grounds Clean/Mainten & -0.0248 & 1.0000 \\ 
  Pers Care/Serv & -0.0287 & 1.0000 \\ 
  Sales & -0.0162 & 0.9980 \\ 
  Office/Administr Supp & -0.0465 & 0.0000 \\ 
  Farm/Fish/Forestry & -0.0498 & 1.0000 \\ 
  Constr & 0.1469 & 0.0810 \\ 
  Install/Mainten/Rep & 0.1496 & 0.0020 \\ 
  Prod & 0.0065 & 1.0000 \\ 
  Transp & 0.0228 & 1.0000 \\ 
  Milit Specific & -0.0799 & 0.9930 \\ 
 [0.8ex]  \multicolumn{3}{l}{\textit{U.S. census region}} \\ [0.8ex]
Middle Atlantic Division & -0.0140 & 0.9980 \\ 
  East North Central Div. & -0.0108 & 1.0000 \\ 
  West North Central Div. & -0.0240 & 0.8980 \\ 
  South Atlantic Division & -0.0117 & 1.0000 \\ 
  East South Central Div. & -0.0374 & 0.2120 \\ 
  West South Central Div. & -0.0346 & 0.0810 \\ 
  Mountain Division & -0.0078 & 1.0000 \\ 
  Pacific Division & -0.0117 & 1.0000 \\ 
\hline
	\end{tabular}
\caption{DOUBLE LASSO results (2/4), bachelor's degree data.}
\label{fullresultsBA2} 
\end{table}

\clearpage

\begin{table}[thb]
\centering 
\begin{tabular}{lrr}
  \hline \\[-0.8ex]
Variable & Estimate & $p$-value  \\ [0.8ex]
\hline \\
		[0.8ex]  \multicolumn{3}{l}{\textit{Metropolitan statistcal area}} \\ [0.8ex]
  msa & 0.0214 & 0.5010 \\  
		[0.8ex]  \multicolumn{3}{l}{\textit{Child}} \\ [0.8ex]
Age $18$ or younger &  -0.0531 & 0.0000 \\ 
  Age $4$ or younger & 0.0809 & 0.0000 \\ 
			[0.8ex]  \multicolumn{3}{l}{\textit{Usual hours worked per week}} \\ [0.8ex]
40 to 49 & -0.0104 & 1.0000 \\ 
  50 to 59 & -0.0048 & 1.0000 \\ 
  60 to 69 & -0.0207 & 0.9980 \\ 
  $>$ 70 & -0.0623 & 0.2150 \\ 
		[0.8ex]  \multicolumn{3}{l}{\textit{Years of education}} \\ [0.8ex]
  yos & 0.0056 & 0.1560 \\ 
		[0.8ex]  \multicolumn{3}{l}{\textit{Experience}} \\ [0.8ex]
  exp & -0.0024 & 0.2770 \\ 
  exp2 & -0.0000 & 0.9270 \\ 
[0.8ex]  \multicolumn{3}{l}{\textit{College major}} \\ [0.8ex]
  Agri & -0.0388 & 0.9640 \\ 
  Envir/Nat Res & -0.0332 & 0.9980 \\ 
  Archit & -0.0316 & 0.9990 \\ 
  Area/Ethnic/Civiliz Stud & -0.0172 & 1.0000 \\ 
  Comm & -0.0133 & 1.0000 \\ 
  Comm Tech & -0.0173 & 1.0000 \\ 
  Comp/Inform Sci & -0.0666 & 0.0000 \\ 
  Cosmet Serv/Culin Arts & 0.1138 & 0.9790 \\ 
  Engin & -0.0545 & 0.0010 \\ 
  Engin Techn & 0.0357 & 1.0000 \\ 
  Ling/Foreign Lang & -0.0267 & 1.0000 \\ 
  Fam/Consum Sci & -0.0322 & 1.0000 \\ 
  Law & -0.0953 & 0.9650 \\ 
  English/Lit/Compos & -0.0140 & 1.0000 \\ 
  Lib Arts/Hum & -0.0330 & 0.9930 \\ 
  Lib Sci & -0.0594 & 1.0000 \\ 
 	\hline
	\end{tabular}
\caption{DOUBLE LASSO results (3/4), bachelor's degree data.}
\label{fullresultsBA3} 
\end{table}

\clearpage
\begin{table}[thb]
\centering 
\begin{tabular}{lrr}
  \hline \\[-0.8ex]
Variable & Estimate & $p$-value  \\ [0.8ex]
  \hline \\ 
[0.8ex]  \multicolumn{3}{l}{\textit{College major (continued)}} \\ [0.8ex]
 Bio/Life Sci & -0.0496 & 0.0040 \\ 
  Math/Stats & -0.0683 & 0.0110 \\ 
  Milit Techn & -0.0554 & 1.0000 \\ 
  Inter-/Multi-Disc Stud (gen) & -0.0851 & 0.1120 \\ 
  Phys Fit/Parks/Recr/Leis & 0.0140 & 1.0000 \\ 
  Philos/Rel Stud & 0.0054 & 1.0000 \\ 
  Theol/Rel Voc & 0.0224 & 1.0000 \\ 
  Phys Sci & -0.0570 & 0.0040 \\ 
  Nucl/Ind Rad/Bio Techn & 0.0834 & 1.0000 \\ 
  Psych & -0.0705 & 0.0000 \\ 
  Crim Just/Fire Prot & -0.0788 & 0.0000 \\ 
  Publ Aff/Policy/Soc Wo & -0.0720 & 0.0670 \\ 
  Soc Sci & -0.0613 & 0.0000 \\ 
  Constr Serv & -0.0982 & 0.9830 \\ 
  Electr/Mech Rep/Techn & -0.1450 & 0.9930 \\ 
  Transp & 0.1077 & 0.9510 \\ 
  Fine Arts & -0.0378 & 0.2050 \\ 
  Med/Hlth Sci Serv & -0.0149 & 1.0000 \\ 
  Bus & -0.0621 & 0.0000 \\ 
  Hist & -0.0561 & 0.0440 \\ 
	\hline
	\end{tabular}
\caption{DOUBLE LASSO results (4/4), bachelor's degree data.}
\label{fullresultsBA4} 
\begin{justify}
 Tables \ref{fullresultsBA1}  to \ref{fullresultsBA4}  present complete results from post-lasso estimation using double selection (DOUBLE LASSO) obtained for the bachelor's degree subsample. $p$-values are obtained from a joint test of all $\beta_j$ coefficients in $\beta(x_i)$ from Equation (3) using the multiplier bootstrap procedure suggested in \cite{BCK-LAD} with 1000 repetitions. 
\end{justify}
\end{table}